\documentclass[aps,prb,twocolumn,showpacs,preprintnumbers,amsmath,amssymb]{revtex4}
\usepackage{graphicx}
\usepackage{color}
\usepackage{dcolumn}
\usepackage{amsmath,amsfonts,mathrsfs,amssymb,amsthm}
\usepackage{hhline}
\usepackage{wasysym,enumerate,color}
\usepackage{epstopdf}
\usepackage{verbatim}
\usepackage{hyperref}

\newcommand{\be}{\begin{equation}}
\newcommand{\ee}{\end{equation}}
\newcommand{\bea}{\begin{eqnarray}}
\newcommand{\eea}{\end{eqnarray}}
\newcommand{\rmd}{{\rm d}}

\newcommand{\al}{{\alpha}}
\newcommand{\tal}{\tilde\alpha}
\newcommand{\ts}{{\tilde s}}
\newcommand{\tsi}{{\tilde \s}}
\newcommand{\bs}{\boldsymbol}

\usepackage{color}
\usepackage{ulem} 
\renewcommand{\emph}[1]{\textit{#1}} 

\definecolor{darkgreen}{rgb}{0,0.5,0}
\definecolor{purple}{rgb}{0.35,0,0.35}
\definecolor{orange}{rgb}{1,0.5,0}
\definecolor{darkred}{rgb}{.7,0,0}
\definecolor{darkblue}{rgb}{0,0,.3}
\definecolor{grey}{rgb}{.6,.6,.6}
\definecolor{dimgreen}{rgb}{0.2,0.6,0.1}
\hypersetup{colorlinks,linkcolor=blue,urlcolor=blue,citecolor=blue}

\newcommand{\e}{\xi}
\newcommand{\sgn}{{\rm sgn}}
\newcommand{\w}{\omega}
\newcommand{\s}{\sigma}
\newcommand{\up}{\uparrow}

\begin{document}

\title{Finite-frequency-dependent noise of a quantum dot in a magnetic field}

\author{C\u at\u alin  Pa\c scu Moca}
\affiliation{BME-MTA Exotic Quantum Phases "Lend\"ulet" Group, Institute of Physics, Budapest University
of Technology and Economics, H-1521 Budapest, Hungary}
\affiliation{Department of Physics, University of Oradea, 410087, Oradea, Romania}
\author{P. Simon}
\affiliation{Laboratoire de Physique des Solides, CNRS UMR-8502, Univ. Paris Sud, 91405 Orsay Cedex, France}
\author{Chung-Hou Chung}
\affiliation{Department of Electrophysics, National Chiao-Tung University,
HsinChu, Taiwan, 300, R.O.C.}
\affiliation{Physics Division, National Center for Theoretical Science, Hsinchu, 30013,
Taiwan, R.O.C.}
\author{G. Zar\'and}
\affiliation{BME-MTA Exotic Quantum Phases "Lend\"ulet" Group, Institute of Physics, Budapest University
of Technology and Economics, H-1521 Budapest, Hungary}

\date{\today}

\begin{abstract}
 We present a detailed study for the finite-frequency current noise  of
 a Kondo quantum dot in the presence of a magnetic field by using a recently 
developed real time functional renormalization group approach [Phys. Rev. B {\bf 83}, 
201303(R) (2011)]. The scaling equations  are modified
in an  external magnetic field; the couplings and non-local 
current vertices become strongly anisotropic, and develop new singularities. 
Consequently, in addition to the natural emission threshold frequency,  $\hbar\omega = |eV|$, 
a corresponding  singular behavior is found to emerge in the noise spectrum at frequencies  
 $\hbar \omega \approx |eV\pm B|$.  The predicted singularities are measurable with present-day experimental
techniques. 
\end{abstract}

\maketitle

\section {Introduction}\label{sec:Introduction}
The study of out-of-equilibrium transport properties of correlated quantum systems is certainly one of the major challenges in condensed matter 
physics. Such correlated  systems often emerge  in mesoscopic physics and molecular electronics, and include, among others, real or artificial atoms and molecules, attached to several electrodes, which are  typically  set to different electrochemical potentials. 
Quantum dots (QDs) -- realizing artificial atoms --  represent the most basic building blocks of these devices. 
Once connected to conduction electrodes, they  behave as artificial impurities interacting with the Fermi sea of conduction electrons on the electrodes attached.   QDs with an odd number of electrons, in particular,   realize artificial magnetic impurities, and thus typically display  a Kondo effect, one of the  most paradigmatic many-body phenomena in condensed matter systems. 

Obviously,  understanding time dependent fluctuations  in such non-equilibrium systems is of major importance. 
In this respect, the noise spectrum of a biased device is a very interesting and important quantity 
since it provides information on the dynamics of excitations on short-time scales.
It is only very recently that  it became possible to investigate  high frequency noise and response functions in mesoscopic circuits in the quantum regime, $\hbar \omega \gg k_B T$ .~\cite{Portier.07,Reulet.08,Portier.10} 
Thanks to progress in on-chip detection of high frequency electronic properties, exploring the nonequilibrium fast dynamics of correlated nanosystems is now accessible, though experiments are delicate since they  involve frequencies in the $ 30\, \div \,100$~GHz range.~\cite{Basset.10} 
In recent experiments, in particular,   high frequency current fluctuations of a carbon nanotube quantum dot
in the Kondo regime have been measured, 
by coupling the QD to a quantum detector via a superconducting resonant circuit.~\cite{Basset.12} 
Strong resonances have been observed in the emission noise for frequencies close to the 
bias voltage,  in agreement with  theoretical predictions.~\cite{Moca.11,Pletyukhov.12}

Compared to experiments, theory is still lagging behind, and describing theoretically how such  many-body states behave under non-equilibrium conditions represents a major unsolved challenge: though there are several approaches to describe electronic transport through correlated mesoscopic circuits,~\cite{Meir.92,Rosch.05,Eckel.10,Konik.01,Kehrein.05,Doyon.06,Anders.08,Pletyukhov.11,Smirnov.13} similarly to the approach presented here, currently none of them is able to describe these correlated states satisfactorily  in all 
regimes of interest. 


In this paper, we shall study the non-equilibrium noise spectrum of a generic strongly correlated mesoscopic element, a quantum dot.
The description of time dependent correlations is particularly challenging in this system, the reason being that the effective interaction between 
a QD and the conduction electrons cannot be treated perturbatively, and an infinite order resummation of the perturbative corrections is needed. 
For a non-equilibrium system, however, this resummation is especially complicated since the effective interaction does not 
only display a singularity at the Fermi energy $\omega\approx 0$, but also exhibits a singular structure whenever the transferred 
energy is in resonance  with the chemical potential difference of the two electrodes, $\omega\approx|\mu_\alpha - \mu_\beta| = |e V_{\alpha\beta}|$.
Simple-minded resummations where the frequency dependence of the effective coupling is neglected, 
cannot account for the aforementioned fine structure, and more sophisticated 
functional renormalization group schemes, similar to the ones used in 
Ref.~\onlinecite{Rosch.01} must be developed. 
Within the latter approach, however, the effective interaction becomes non-local in time, 
and the definition of instantaneous current operators satisfying current conservation --- and thus the computation of time dependent correlation functions --- becomes a particularly 
delicate issue.~\cite{Moca.11}  It is probably for this reason that most calculations focused so far on the zero frequency (shot noise) 
limit~\cite{Meir.02, Thielmann.03, Sela.06, Gogolin.06, Herrmann.07,Yamauchi.11}, and finite frequency results  are rather 
limited.~\cite{Engel.04,Braun.06,Schiller.98,Moca.11,Basset.12,Fritsch.09,Schuricht.09,Fritsch.10}

In  Ref.~\onlinecite{Moca.11} we developed a current conserving real time functional renormalization group (RG) formalism, and have 
shown that the somewhat intuitively derived  equations of Ref.~\onlinecite{Rosch.03} follow easily within this formalism. 
We have shown that, similar to the effective interaction (vertex),   local measurables (in time)  also develop a 
non-local character in the course of the  RG process. We then used this formalism to compute the noise spectrum of a 
quantum dot in the local moment regime,~\cite{Moca.11} and compared that to the experimentally measured 
noise spectrum,  to find good agreement.~\cite{Basset.12}
Notice that our real time FRG approach is different from the FRG approach used by Metzner \emph{et al.}\cite{Metzner.12} and 
 Kopietz \emph{et al.}~\cite{Kopietz.01}: similar to the real time RG of 
Schoeller and K\" onig,\cite{Schoeller.00}  we work on the Keldysh contour, however, unlike the usual 
FRG method we perform the RG transformation on the bare action, and keep only quartic but nonlocal interaction terms.

 \begin{figure}[t]
\includegraphics[width=1\columnwidth,clip]{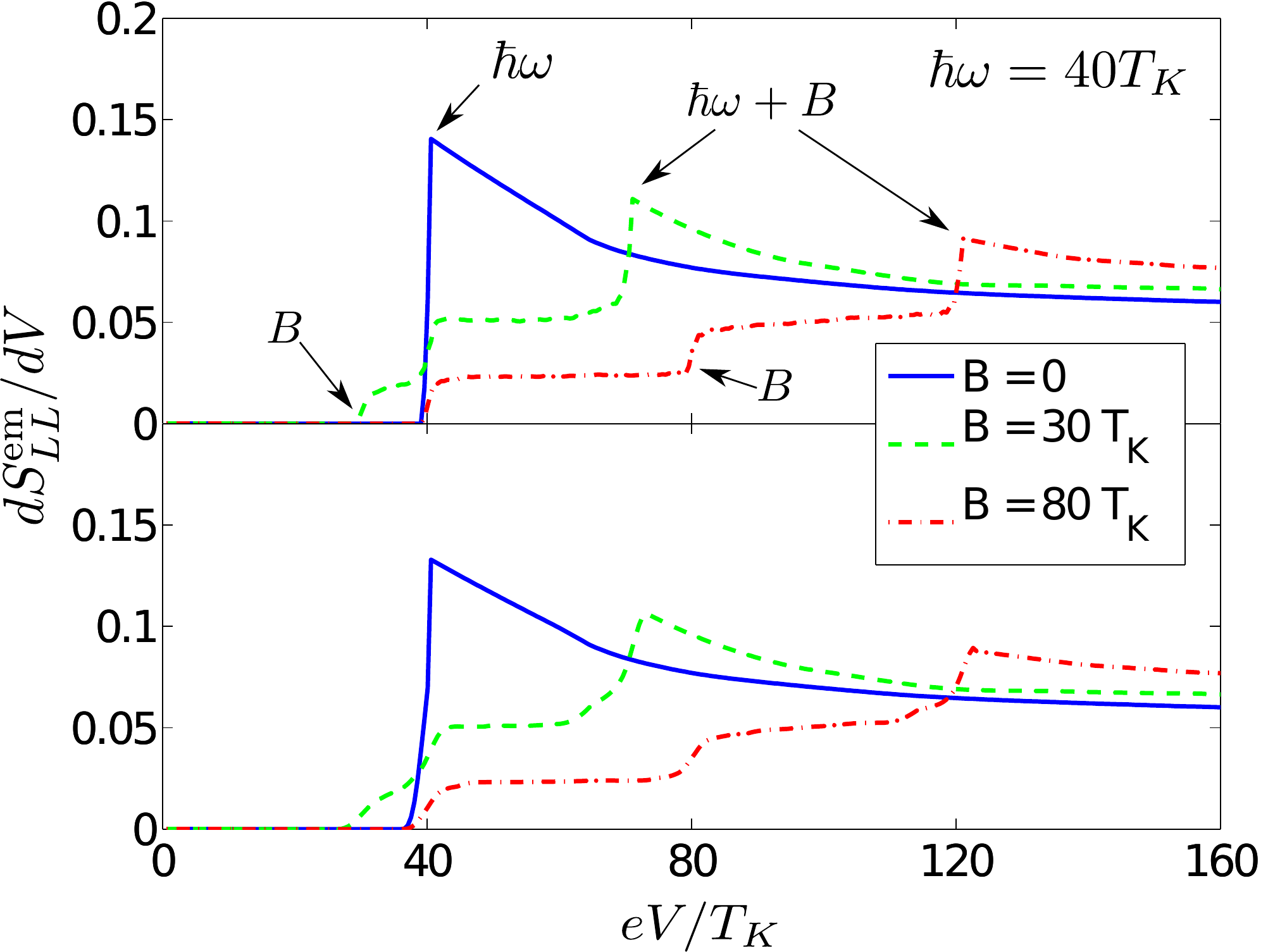}
\caption{ (color online)
Voltage dependence of the differential emission noise, $\mathrm{d}S^{e}_{LL}(V, T=0)/\mathrm{d}V$ at a constant  
frequency, $\omega=40\; T_K$ for various Zeeman splittings $B$. All energies are measured in units of the Kondo temperature, 
$T_K$. In the upper figure, the spin decay time has only been incorporated when computing the  current vertex renormalization, while in the lower panel a finite pseudofermion lifetime has been used to compute the diagrams in Fig.~\ref{fig:noise}.
}
 \label{fig:emission_noise_derivative}
\end{figure}
 
Here we intend to give a more detailed description of the formalism presented in Ref.~\onlinecite{Moca.11}, and 
extend it to the case where an external field is also present. Throughout this paper, we shall focus most of our attention 
on the Kondo model, where the external magnetic field $B$ couples to  
 the impurity spin operator, $\mathbf{S}$,\footnote{We consider the effect of a local Zeeman field. A field applied to the 
 conduction electrons  has a similar effect.} 
$$
H_B = -B S^z\,,
$$
and  the (unrenormalized) interaction is of  a simple exchange form, 
\begin{equation}
H_{\rm int}=  \frac 1 2 \sum_{\alpha,\beta\in L,R} \sum_{\sigma,\sigma'}
j_{\alpha\beta}\;
\;\psi_{\alpha\sigma}^\dagger \mathbf{S} \cdot {\bs \sigma}_{\sigma\sigma'}\psi_{\beta\sigma'} \;,
\label{eq:H_int}
\end{equation}
with  $\psi_{\alpha\sigma}^\dagger $ the creation operator of an electron 
of spin $\sigma$ in lead $\alpha\in\{L,R\}$, 
$\bs \sigma$ the Pauli matrices,
and the $j_{\alpha\beta}$ denoting dimensionless exchange couplings. 
Nevertheless, our formalism  is very general, and can be applied to any local 
quantum impurity problem with a "quantum impurity" having some internal quantum states, 
$s\in\{1,\dots, Q\}$ of energy $E_s$ and 
 interacting with the leads through the Hamiltonian
\begin{eqnarray}
H_{\rm int} &=&   \sum_{i,k } \sum_{s,s'}
g^{ss'}_{ik}\;   |s \rangle\langle s'|
\;\psi_{i}^\dagger \psi_{k}\;,
\label{eq:H_int_gen}
\end{eqnarray}
with  $i$ and $k$ labeling conduction 
electron channels of different chemical potentials, $\mu_i$ and $\mu_k$, respectively.\footnote{In the Kondo case, $i$ is a composite label standing simultaneously for spin and lead indices $i=(\alpha,\sigma)$.}
The fields $\psi_{\alpha\sigma} $ (and $ \psi_i$) in the previous equations are constructed in terms of quasiparticle operators, 
\be 
\psi_{\alpha\sigma} = \int  c_{\alpha\s}(\e) e^{-|\e| a/2}\;d\e,
\label{eq:psi}
\ee
 with $a$ a short time ($1/a$ a high energy) cut-off, and 
the  operators $c_{\alpha\sigma}(\xi)$ destroying a quasiparticle of energy $\xi + \mu_\alpha$, in lead 
 $\alpha$ (of energy $\xi + \mu_i$ in channel $i$ in the general case).Ê\footnote{The operators $c_{\alpha\sigma}(\xi)$ are normalized such that 
$\{c_{\alpha\sigma}(\xi), c^{\dagger}_{\alpha'\sigma'}(\xi') \} = 
\delta_{\alpha\alpha'}\delta_{\sigma\sigma'}\delta(\xi-\xi')$.}

The non-equilibrium quantum impurity problems defined by Eqs.~\eqref{eq:H_int} and \eqref{eq:H_int_gen} constitute 'hard problems', and do not possess  complete solutions yet. The Kondo problem, Eq.~\eqref{eq:H_int} has, however, an exact solution under equilibrium conditions,~\cite{Andrei.83,Tsvelick.83} and is well-understood.~\cite{Wilson.83,Affleck.93,Hewson.93} For a quantum dot, in particular, the couplings $j_{\alpha\beta}$ assume the simple form,  $j_{\alpha\beta} = j\;  v^*_\alpha v_\beta$, with $v_\alpha$ a complex two-component spinor of unit length. The dimensionless coupling $j$  generates a dynamical energy scale, the so-called Kondo temperature\footnote{The prefactor in the expression of $T_K$ is not universal, and depends on  the exponential cut-off scheme in Eq.~\eqref{eq:psi}.  We determined it from the RG equations by identifying the energy at which the renormalized couplings diverge.}
$$
T_K \approx \frac {1.4}{a} \, e^{-1/j}\;. 
$$
Below this energy scale the effective exchange coupling becomes infinitely strong, and a local Kondo singlet is formed.~\cite{Wilson.83}
Similarly, in equilibrium, the effective couplings  $g^{ss'}_{ik}$ of the generalized problem,  Eq.~\eqref{eq:H_int_gen} would typically scale to strong coupling  below some Kondo scale provided that some of the levels $E_s$ are degenerate. 
 
Here we shall not attempt to reach this strong coupling regime,~\cite{Pletyukhov.12} rather, we restrict ourself to the weak coupling  regime, where either the voltage or the temperature, or the Zeeman splitting is sufficiently large compared to $T_K$.  
Our  main goal  is  to investigate in detail the properties of the correlation functions of the current operators in this so-called 'weak coupling' regime, 
and to determine the frequency dependent conductance of a biased device as well as its emission/absorption
noise spectrum, and the symmetrized noise, accessible through direct measurements of 
the noise spectrum.~\cite{Reulet.08} 
Particular attention shall be devoted to spin relaxation processes, which are crucial to provide a self-consistent cut-off to the 
singular structures.    As we shall see, in a two terminal Kondo device, all noise components  exhibit interesting, singular  structures at the thresholds  $\omega= \pm eV$ and $\omega= \pm |eV\pm B|$.  The
differential emission noise spectrum $\mathrm{d}S^{\rm em}(\omega)/\mathrm{d}V$, 
in particular, exhibits a peak at $\omega\approx eV$,~\cite{Moca.11,Basset.12} 
which is then split in a magnetic field, as shown in 
Fig.~\ref{fig:emission_noise_derivative},
in agreement with the recent independent results of Ref.~\onlinecite{Muller.13}.
Similar structures are predicted  in the symmetrized noise of the device (see Figs.~\ref{fig:symmetric_noise_frequency}).

The paper is organized as follows: In Sec.~\ref{sec:real_time_approach} we introduce 
the real time functional renormalization group formalism, with
special emphasis on the construction of the Keldysh action 
 and the  derivation the scaling equations for the couplings. The 
current operator and the corresponding equations for the current vertices are discussed
in Sec.~\ref{sec:current_scaling}. In Sec.~\ref{sec:decoherence} we discuss the issue of 
decoherence in terms of the Master Equation approach and present results for the 
pseudofermion self-energy, while the result for the finite frequency noise are presented in Sec.~\ref{sec:noise}. We give the 
final remarks in Sec.~\ref{sec:conclusions}.


\section{The real-time functional renormalization group approach}
\label{sec:real_time_approach}

In this section, we shall present in detail the real-time functional renormalization group (RTFRG) we have developed in Refs.~\onlinecite{Basset.12}
 and~\onlinecite{Moca.11}. First, we discuss how our functional RG formalism is  constructed on the Keldysh contour, and show 
how the RG equations of Ref.~\onlinecite{Rosch.05} can be derived within this formalism. Then, in the next section
 we  discuss how the RG equations can be extended to the current vertex (subsection~\ref{sec:current_vertex}).

\subsection{Keldysh action}\label{sec:Keldysh}

For the non-equilibrium physics discussed here it is most convenient to work with a path integral formalism on the Keldysh contour. 
This approach allows one to incorporate retardation effects in a natural way, and does not suffer from the 
restrictions of a Hamiltonian theory, where the renormalized theory is  local. 
In this approach fermionic fields 
are replaced by time-dependent Grassmann fields living 
on the Keldysh contour, $\psi^\dagger_{\alpha\s}(t)\to \overline  \psi_{\alpha\s}(z)$, ($z\in K$).
As usual,  the branches  $z\to t_\pm =  t \pm  i\delta $ of the contour are labeled by a Keldysh label $\kappa=1,2$, and they 
represent forward/backward propagations in time. 

 The dynamics of the systems is determined  by the Keldysh action,
\begin{equation} 
{\cal S} = {\cal S}_{\rm lead} +{\cal S}_{\rm spin} +{\cal S}_{\rm int}. 
\end{equation}
The terms ${\cal S}_{\rm lead}$ and ${\cal S}_{\rm spin}$ describe the 
 conduction electrons action and the spin action in the absence of interaction. They
are quadratic in the fields and determine  the non-interacting Green's functions (see below).
  
The part   ${\cal S}_{\rm spin}$ describes the spin, which we represent  using Abrikosov's 
pseudofermions~\cite{Abrikosov.65} as 
$S^i \to \frac 12 \sum_{s,s'} f^\dagger _{s} { \sigma}^i_{ss'} f_{s'}$, with the 
pseudofermion operators $f^\dagger _{s}$ satisfying the constraint 
$\sum_{s} f^\dagger _{s}  f_{s} \equiv 1$. Correspondingly, in the path integral language the spin part of the Keldysh 
action simply reads 
\be
{\cal S}_{\rm spin} =   \int_{z \in K} \mathrm dz\sum_{s} \overline f_s(z) (  - i \partial_z +  \lambda_s) f_{s} (z)\,, 
\label{S_spin}
\ee
which can also be expressed in terms of the Keldysh labels $\kappa$ as 
\be
{\cal S}_{\rm spin} =  \sum_{\kappa} \int_{-\infty}^\infty  \mathrm dt \sum_{s} \overline f^{(\kappa)}_s(t) (  - i \partial_t + s_\kappa  \lambda_s) f^{(\kappa)}_{s} (t)\,, 
\label{eq:S_spin2}
\ee
with the sign $s_\kappa$ being $s_\kappa =\pm$ for the upper ($\kappa=1$) and lower ($\kappa=2$) contours.
Here the chemical potentials $\lambda_s = \lambda_0 + s B/2$ account for  the splitting of the two spin states, $s=\pm$,  
but they also act as Lagrange multipliers to implement the constraints, and allow us to separate in the $\lambda_0\to\infty$ limit 
the contribution of the states 
satisfying  $\sum_{s} f^\dagger _{s}  f_{s} \equiv 1$. 
The actions  \eqref{S_spin} and \eqref{eq:S_spin2} determine the four pseudofermion correlation functions 
$F^{t / \overline t}_s $ and $F^{\gtrless}_s $ depending on the choice of the Keldysh branches $\kappa$ and $\kappa'$ (see Appendix~\ref{app:Greens_functions}).
The time ordered propagator, e.g., is given by the $'11'$ component, $ F_s^{t}(t)= F^{(11)}_s (t) $,
\bea
F^{(11)}_s (t) &=&-i\langle f^{(1)}_{s}(t) \overline f^{(1)}_{s}(0)\rangle_{\cal S_{\mathrm{spin}}}
=
-i\langle T_t f_{s}(t) f^\dagger_{s}(0)\rangle\
\nonumber 
\\
&\approx&
 -i \,e^{-i\lambda_s t}\bigl ( \Theta(t) - e^{-\beta \lambda_s} \overline \Theta(t)\bigr)\,,
\eea
with $T_t$ denoting the time ordering and $\Theta(t)$ and $\overline \Theta(t)= 1 - \Theta(t)$ the forward and backward 
step functions, respectively.

The interaction part of the action, ${\cal S}_{\rm int}$, is initially local in time. 
However, as we shall see,
the elimination of high energy degrees of freedom -- implemented in our scheme by rescaling $a$ in Eq.~\eqref{eq:psi} --  generates retardation effects
 in the course of the  RG procedure
and the interaction becomes therefore {\it non-local} in time.
We therefore replace the Kondo couplings by some
time-dependent vertex functions, $g^{\sigma s;\sigma's'}_{\alpha\alpha'}(t)$, which depend on the incoming and outgoing
electrons' spin and lead  indices and on the pseudo-fermion spins, and  write the interacting part  of the action  as
\bea 
{\cal S}_{\rm int} &=&
\sum_\kappa \sum_{\alpha\alpha' \sigma\sigma'}  \sum_{s,s'}
 \int {\rm d}t\;{\rm d}t'\;
 s_{\kappa}\;
 g^{\sigma s;\sigma' s'}_{\alpha\alpha'}(t-t')\;
\nonumber
\\
&& \bar f^{(\kappa)}_{s}(\bar t)  f^{(\kappa)}_{s'}(\bar t)  \cdot
\bar \psi^{(\kappa)}_{\alpha\sigma}(t)  \psi^{(\kappa)}_{\alpha' \sigma'}(t') \;. 
\label{eq:S_int}
\eea
Here $\bar t= (t + t')/2$, and the tensor  function $g^{\sigma s;\sigma' s'}_{\alpha\beta}(t)$ collects the various components of 
 the Kondo couplings.
Initially,  the couplings  $g^{\sigma s;\sigma' s' }_{\alpha\beta}(t)$ are local in time, and are given by
\be
g^{\sigma s;\sigma's'}_{\alpha\alpha'}(t)
= \frac {j_{\alpha\alpha'}} 4 \;{\bs \sigma}_{ss'}\cdot \bs\sigma_{\sigma\sigma'}\;
\delta(t)\,.
\label{eq:g_initial}
\ee
In the absence of an external field, this SU(2) invariant structure is conserved, 
$g^{\sigma s;\sigma's'}_{\alpha\alpha'}(t) \to 
( {g_{\alpha\alpha'}(t)}/ 4 )\;{\bs \sigma}_{ss'}\cdot \bs\sigma_{\sigma\sigma'}$, 
and it is enough to keep track of  just four functions, $ g_{\alpha\alpha'}(t)$ 
[see also Eq. \eqref{eq:full_couplings}].

\begin{figure}[t]
\includegraphics[width=7cm]
{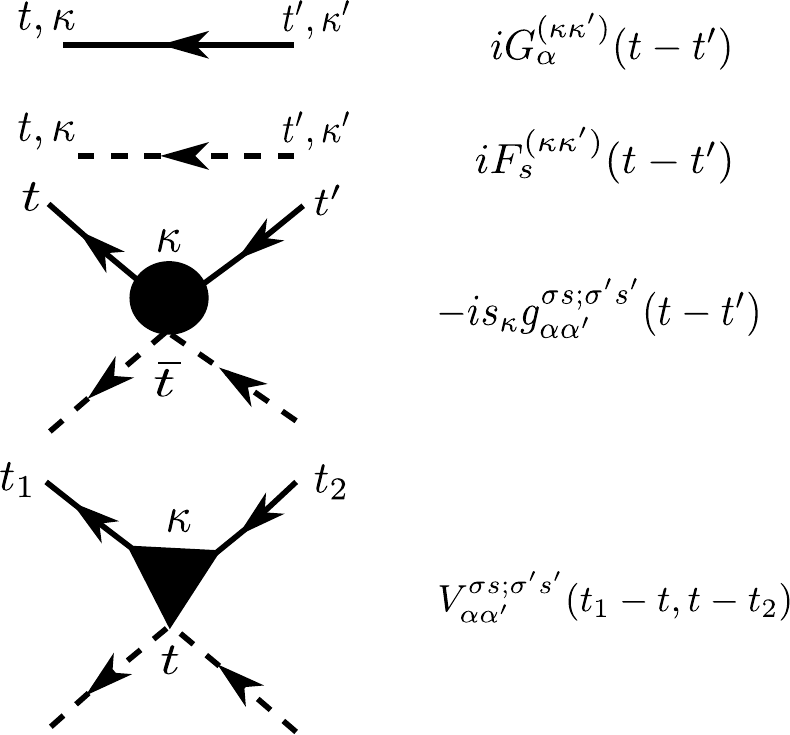}
\caption{\label{fig:diagram_components} 
Components of  real time Feynman diagrams. Electron and pseudofermion propagators are denoted by continuous and dashed lines, respectively. Incoming and outgoing electrons carry  Keldysh ($\kappa$), spin ($\sigma$) and 
lead ($\alpha$) quantum numbers, while pseudofermions carry only Keldysh and spin ($s$) labels.  
The non-local vertex $g^{\sigma s; \s' s'}_{\alpha\beta}(t)$ is indicated by a filled circle. The time argument of the pseudofermion 
is  $\bar t=(t+t')/2$. Finally, the current vertex is labeled by three time arguments: the time $t$ of the measurement, 
and the times $t_1$ and $t_2$ of the incoming and outgoing electrons.}
\end{figure}

The structure in Eq.~\eqref{eq:S_int} can be justified by observing that the spin evolves 
very slowly at electronic time scales and interacts only weakly with the electrons. Therefore  
its time evolution can be well approximated by that of a free spin (pseudofermion). In contrast, conduction electrons 
have fast dynamics, and their  scattering on the slow impurity spin induces retardation 
effects which become more and more pronounced 
as one approaches smaller and smaller  energy scales or, equivalently, long time scales.
These are precisely the effects we  want to capture within our formalism. Technically, this implies that 
we need to keep the time arguments of the electron fields $\overline \psi_{\alpha\sigma}(t)$ and 
$\psi_{\alpha'\sigma'}(t')$, while we can eliminate the time evolution of the pseudofermion fields
using their bare real time evolution, $f_s(t)\approx e^{-i \lambda_s (t-t')}f_s (t')$ for short time differences.
From a diagrammatic point of view, we can represent the interaction term 
\eqref{eq:S_int} by a non-local vertex diagram, depicted in Fig. \ref{fig:diagram_components}.  

Notice  that 
 the couplings in \eqref{eq:S_int} do not have a Keldysh label, and that all fields live  on the same branch of the 
 Keldysh contour. This is obvious for the bare action, but  this structure is also  approximately maintained by the 
  renormalized action as long as only singular terms are summed up (see Section~\ref{sub:RG}).

Finally, the term $\cal S_\mathrm{lead}$ describes the electrons in the leads and generates the  non-interacting Keldysh Green's functions
of the fields $\psi_{\alpha\sigma}(t)$. 
At $T=0$ temperature, e.g., a simple calculation yields for the time ordered and greater propagators
\bea
G_{\alpha\s;\alpha'\s' }^t(t)&=&-i\langle T_t \psi_{\alpha\sigma}(t)\psi^{\dagger}_{\alpha'\sigma'}(0)\rangle
\nonumber 
\\
&=& \delta_{\alpha\alpha'} \delta_{\sigma\sigma'} G_{\alpha }^t(t) = 
 \frac{-  \delta_{\alpha\alpha'} \delta_{\sigma\sigma'} e^{-i \mu_\alpha t} }  
{t-i a\; {\rm sgn}(t)}
\nonumber
\,,
\\
G_{\alpha\s;\alpha'\s' }^>(t)&=&-i\langle \psi_{\alpha\sigma}(t)\psi^\dagger_{\alpha'\sigma'}(0)\rangle
\nonumber 
\\
&=& \delta_{\alpha\alpha'} \delta_{\sigma\sigma'} G_{\alpha }^{> }(t)
=\frac{- \delta_{\alpha\alpha'}  \delta_{\sigma\sigma'}  e^{-i \mu_\alpha t}}  {t-i a} 
\nonumber \, .
\eea
The other electronic propagators are given by similar expressions and are listed 
 in Appendix~\ref{app:Greens_functions}.

\subsection{ Derivation of the RG equations for the couplings}\label{sec:scaling_g}
\label{sub:RG} 

Let us now turn to the derivation of the RG equations. In this section 
we shall proceed  by using Wilson's RG  approach within a path integral 
formalism:~\cite{Cardy.96} 
we gradually increase the time scale $a\to a'$ thereby eliminating high energy electronic degrees of freedom, 
and compensate for the reduction  of the cut-off $1/a\to 1/a'$ by renormalizing the vertex function ${\bs g}(t)\to {\bs g}'(t)$. 

\begin{figure}[t]
\includegraphics[width=7cm]
{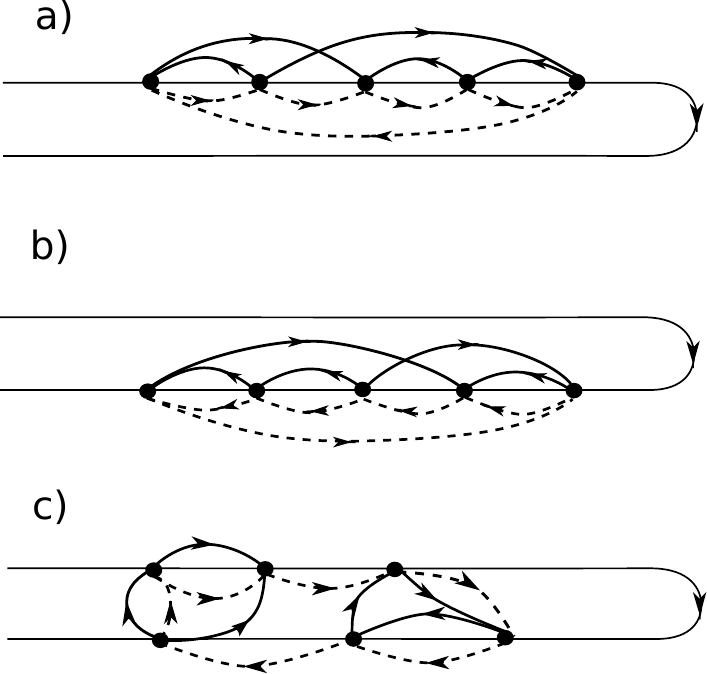}
\caption{\label{time_ordered_diagrams} 
The physical subspace  $\sum_sf^\dagger _s f_s =1$  corresponds to 
diagrams with a single time-ordered  pseudofermion loop. }
\end{figure}
 
To determine the renormalization of ${\bs g}(t)$, let us assume an interaction vertex of the 
form \eqref{eq:S_int} and expand the functional 
$$
{\cal Z} \equiv \int {\cal D}f{\cal D}\overline f \int {\cal D}\psi{\cal D}\overline \psi\; e^{-i \cal S}
$$
in ${\cal S}_\mathrm{int}$. 
The contributions 
of $n$'th order  diagrams can be evaluated using Wick's theorem, and can be represented by Feynman diagrams.
The diagrammatic components are shown in Fig.~\ref{fig:diagram_components}.

It is relatively easy to see that, as a result of the structure of the pseudofermion Green's functions 
listed in Appendix~\ref{app:Greens_functions}, each pseudofermion loop contains at least 
one exponentially small pseudofermion propagator, $\sim e^{-\beta \lambda_s}$, since for any time configuration it involves 
at least one of the three propagators, $F^{<}_s(t)$, $F^{t}_s(t<0)$, or $F^{\bar t}_s(t>0)$. 
The physical subspace, however, corresponds to having {\it exactly} one pseudofermion, which has a probability 
$P_1 = \sum_s e^{-\beta \lambda_s}$. In fact, when computing physical quantities,  the contribution of 
every diagram must be normalized by this probability, while the chemical potential of the pseudofermions is taken to infinity, 
$\lambda_s\to\infty$.  Thus the physical subspace $\sum_sf^\dagger _s f_s =1$ corresponds to 
diagrams with a single time-ordered  pseudofermion loop.

It follows by the same argument   that, upon integration over all time arguments of a given diagram, only those 
 time configurations give a contribution where the pseudofermion lines are time ordered along the Keldysh loop
 (see Fig.~\ref{time_ordered_diagrams}).  The pseudofermion fields thus lead to an effective time ordering along the Keldysh contour.  The contribution of all diagrams  in  Fig.~\ref{time_ordered_diagrams}  is thus proportional to 
$\sim e^{-\beta \lambda_s}$, which is, as explained above, proportional to the probability of having exactly one pseudofermion available. 
Of course, time ordering is automatically performed  by the $\Theta$ functions in the pseudofermion propagators, upon integration over all internal time arguments of a given diagram.

Let us now investigate the effect of changing $a\to a' = a + \delta a$.  In a given $n$'th order  diagram we need to perform 
$2n$ integrations over some times on the upper and on the lower contour.  However, under the rescaling 
 $a\to a'$  the value of the integrand changes substantially only
when  two of the  contracted time arguments of the fields $\psi_{\alpha\sigma}^{(\kappa)}$, say $t_1$ and $t_2$ happen to be close
to each other, $|t_{12}| \equiv  |t_1-t_2|\sim a$, implying that all time arguments of the two corresponding  vertices must  also be close to each other. This is obvious from the structure of the fermionic Green's functions, which change 
only locally upon rescaling  $a\to a'$. The change in $G_\alpha^t(t_1-t_2)$ is, e.g., approximately equal to
$$
\delta G_\alpha^t(t_{12})\approx   -i  e^{-i \mu_\alpha t_{12}} \frac{ \delta  a\; {\rm sgn}(t_{12})}  
{(t_{12}- i a\; {\rm sgn}(t_{12}))^2}
$$
at $T=0$ temperature, and decays asymptotically  as ${\rm sgn}(t_{12}) /t_{12}^2$. Similarly, the rescaling of any
other electronic Green's function  gives a short range contribution in time. 
Therefore,  we can safely assume that  the typical distance of $t_1\approx t_2$ from the time arguments 
of all other  vertices  is large compared to 
$|t_1-t_2|\sim a$. Under this assumption,   we can integrate over 
contracted  local time arguments,  $t_{1}$ and $t_2$, and compensate the change  $\delta G^{(\kappa\kappa')}_{\alpha}$ by
adding a corresponding  interaction term to the action.

\begin{figure}[t]
\includegraphics[width=0.9\columnwidth, clip]{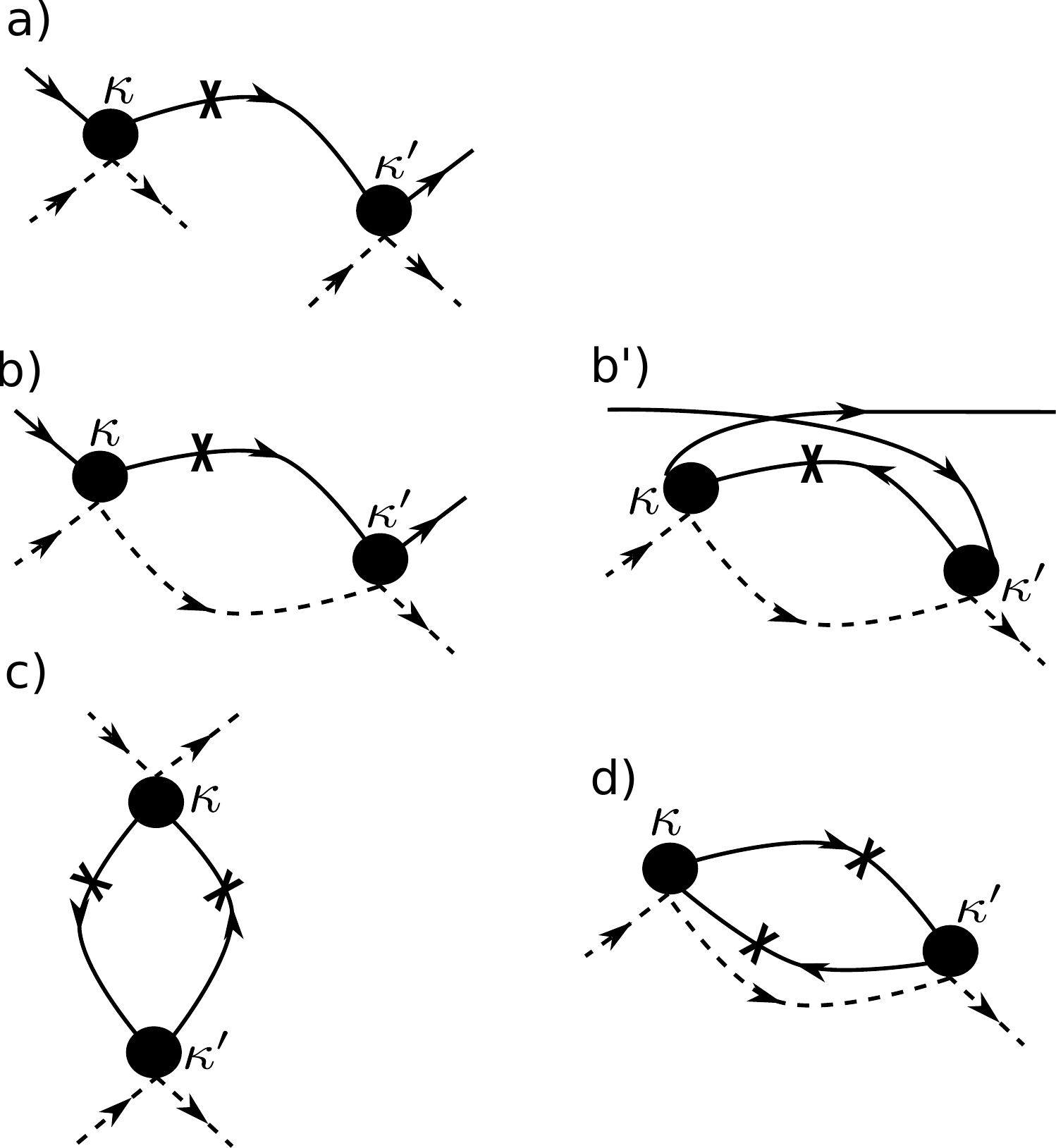}
\caption{\label{fig:contractions} 
Possible contractions of two vertices with time arguments at a distance $\sim a$. Contractions of type b) and b') renormalize the 
interaction, while contractions of type d) renormalize the pseudofermion propagators and account for spin relaxation.
Crosses indicate the logarithmic derivative with respect
to the scaling parameter $a$.}
\end{figure}

\begin{figure}[tbh]
\includegraphics[width=0.9\columnwidth,clip] {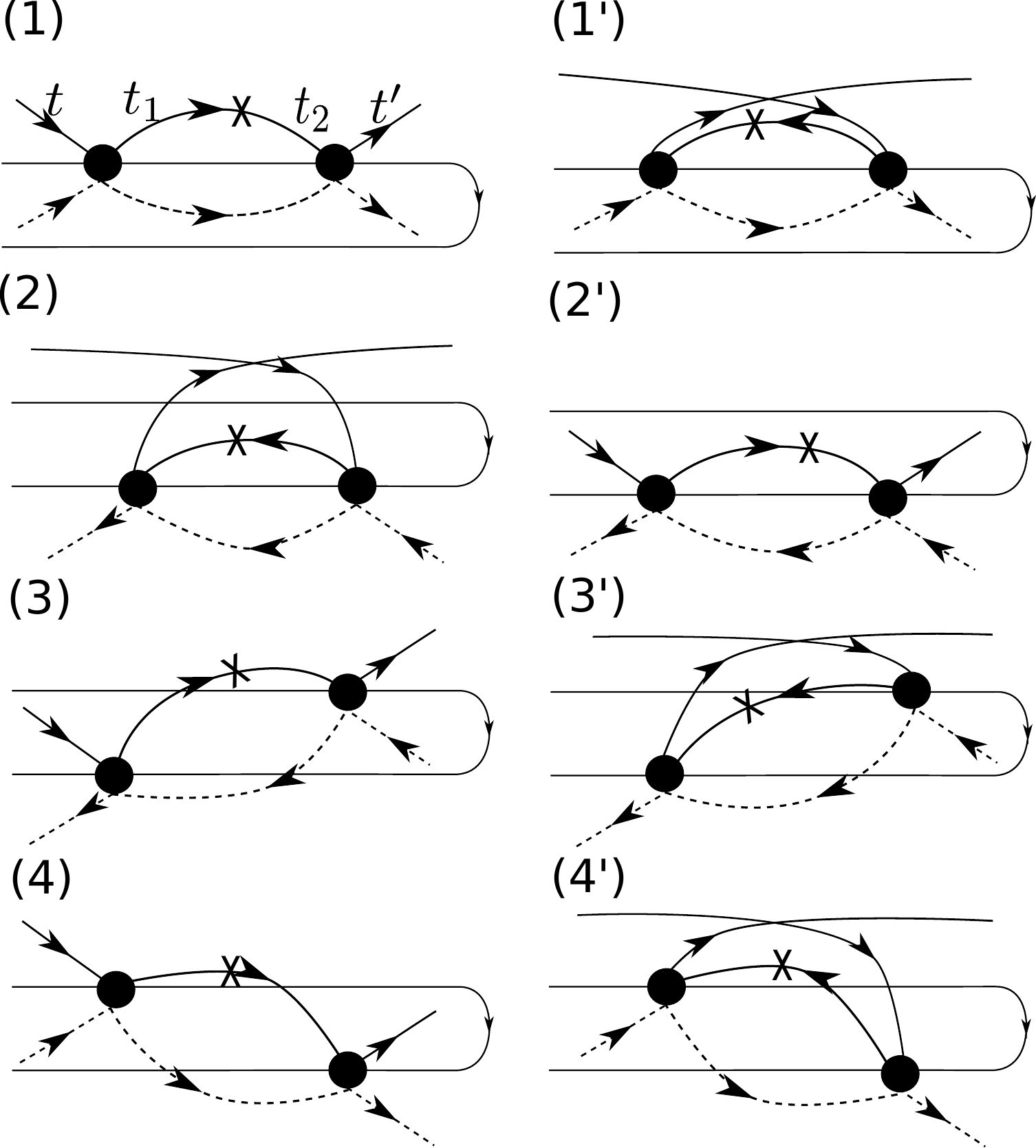}
\caption{\label{fig:couplings_diagram} 
Possible vertex corrections. To leading order, only diagrams (1),  (1'), (2),  and (2') contribute to vertex renormalization.Crosses indicate the logarithmic derivative with respect
to the scaling parameter $a$.}
\end{figure}

Let us now focus on two vertices as being part of a big diagram containing $n>2$ vertices. Using Wick's theorem, we can 
write the contribution of these two vertices as a sum of normal ordered operators (which contain fields $f$, $\overline f$,
$\psi$, $\overline \psi$ to be contracted with external vertices) multiplied by certain internal contractions.  
Typical contractions are sketched  in Fig.~\ref{fig:contractions}. We only show those  contractions which contain at least 
one $\psi$-contraction, since only these contributions change upon rescaling $a$, and therefore only these diagrams can give a 
contribution to the renormalized action, at least in leading order in $a$. Furthermore, as a basic principle, we shall keep only those diagrams which do not vanish  in the equilibrium limit,  $B\to 0$ and $\mu_L = \mu_R$. We have the following four classes: 
\vspace{0.2cm}
\\
(a) Diagrams with a single $\psi$-contraction. Such diagrams do not give a contribution for the following reasons: If one of the vertices is 
on contour $\kappa=1$  and the other on contour $\kappa'=2$, then its change is proportional to $\delta G^{\gtrless}(t_1-t_2)\sim 1/(t_{12}\pm i\,a)^2$. The strength of this correction can be estimated within the local approximation, ${\bs g}(t)\to {\bs g}\,\delta(t)$, whereby one replaces the fields $\bar \psi(t)\to \bar \psi(\bar t)$ and $ \psi(t')\to \psi(\bar t)$, and integrates over the internal variable, $t_{12}$. This procedure 
 yields  a vanishing contribution.  If, on the other hand, the two vertices are on the same 
Keldysh branch, then at least one  pseudofermion leg of the two vertices must be connected.  
This follows from the observation that,  to give a non-vanishing (i.e. $\sim e^{-\beta\lambda_0}$) contribution,
the time arguments of the pseudofermions must be contracted to form an ordered loop along the contour.
Diagrams of type (a) can thus be ignored. 
\vspace{0.2cm}
\\
(b) and (b') Diagrams with one pseudofermion line contraction and one $\psi$-contraction. These diagrams
renormalize $\bs g$, as shall be discussed in detail below. 
\vspace{0.2cm}
\\Ê
(c) Diagrams with two $\psi$-contractions. These diagrams account for the relaxation of the spin's density matrix, and 
incorporate information on the Korringa relaxation.  We shall neglect these diagrams, and only keep track of the spin relaxation 
through the pseudofermion's relaxation [diagram (d)]. 
\vspace{0.2cm}
\\Ê
(d) Diagrams with two $\psi$-contractions and one pseudofermion contraction. These diagrams generate a pseudofermion 
self-energy, and account for (at least part of the) spin relaxation.

Let us now focus on the vertex renormalization, i.e., on the family of diagrams (b) and (b') in Fig.~\ref{fig:contractions}.
Depending on the Keldysh labels of the two vertices, these give rise to 4 + 4 diagrams, as shown in Fig.~\ref{fig:couplings_diagram}. 
As an example, let us 
 discuss  the first diagram (diagram (1) in  Fig. \ref{fig:couplings_diagram}), on the upper Keldysh  contour. 
By rescaling $a$, we generate the following  term in the effective action,

\begin{widetext}
\begin{multline}
-i {\cal S}_{\rm int}^{(1)}
= 
 \int {\rm d}t\, {\rm d}t'\,
\sum_{\al,\al', \s,\s'}
\sum_{ s,s'}
\left (
\sum_{\tal,\ts,\tsi}
\int {\rm d}t_1\, {\rm d}t_2\,
g_{\al\tal}^{\s\tsi ;  s\ts}(t-t_1)g_{\tal\al'}^{\tsi \s'; \ts s'}(t_2-t')
\right . \times
\\
\left .
 \delta G^{t}_{\tal}(t_1-t_2) \, F^{t}_{\ts}\left(\frac{1}{2} (t+t_1-t_2-t' )\right)
\,e^{i\lambda_s(t_1-t')/2}
\, e^{-i\lambda_{s'}(t_2-t)/2}
\right )
 \bar f^{(1)}_{s}(\bar t)  f^{(1)}_{s'}(\bar t)  \cdot
\bar \psi^{(1)}_{\alpha\sigma}(t)  \psi^{(1)}_{\alpha' \sigma'}(t') \;,
\label{eq:delta_H}
\end{multline}
\end{widetext}
where $\bar t=(t+t')/2$.
In order to obtain this equation, we have used the  expansion 
\bea
&\bar f_s^{(1)}(\frac{t+t_1}{2})\approx \bar f^{(1)}_s(\bar t)\,e^{i\lambda_s(t_1-t')/2}\;\;\;{\rm and}
\nonumber
\\
&f^{(1)}_{s'}(\frac{t_2+t'}{2})\approx f^{(1)} _{s'}(\bar t)\, e^{-i\lambda_{s'}(t_2-t)/2},
\nonumber
\eea
which assumes a slow spin dynamics compared to that of the electrons. 
We thus conclude that we can compensate for the change of the Green's function 
$\delta G^{t}_\al$ in diagrams of type (1) in Fig.~\ref{fig:couplings_diagram}
by renormalizing  the interaction kernel $g(t)$ on the upper Keldysh contour by
\begin{multline}
\delta g^{(1)} (t-t') = 
\sum_{\tal,\ts,\tsi}
\int {\rm dt_1\, dt_2}\,
g_{\al\tal}^{\s s;\tsi \ts}(t-t_1)g_{\tal\al'}^{\tsi \ts;\s' s'}(t_2-t')
\\
\delta G^{t}_{\tal}(t_1-t_2) F^{t}_{\ts}\left(\frac{1}{2} (t+t_1-t_2-t' )\right)
\\
\exp\left \{i\frac{\lambda_s}{2}(t_1-t')\right \}
\exp\left \{-i\frac{\lambda_{s'}}{ 2}(t_2-t)\right \}.
\label{eq:deltag^(1)}
\end{multline}
The contributions of the other diagrams can be treated similarly. However, while diagrams 
(1), (1'), (2) and (2') lead to  changes of time- and anti- time ordered  electronic propagators $\sim \sgn \,t_{12}/(t_{12}\pm ia \,\sgn \,t_{12})^2 $, integrating 
to a finite value $\sim 1/a$,  changing the electron propagators in diagrams (3), (3'), (4) and (4') results in 
  terms $\sim 1/(t_{12}\pm ia)^2 $, and  integrate to $\approx 0$. Notice that in the latter four diagrams the pseudofermion propagators 
 do not contain $\Theta$ functions. Therefore, these diagrams do not result in any interesting renormalization.  
 Put in another way, the parent diagrams of (1), (1'), (2) and (2') (without the crosses) contain
\emph{logarithmic singularities} associated with the contraction of $t_1$ and $t_2$, while the diagrams (3-4') contain no such singularity. 
In the spirit of leading logarithmic approximation, where only maximally singular diagrams are kept, 
we thus drop the latter two sets of diagrams. Notice that, within this approximation, 
 the generated vertex functions do \emph{not have} off-diagonal  Keldysh labels. Furthermore, one can show that   
the contributions of diagrams (3) and (4) are identical to those of (1) and (2), and therefore the  structure 
of  Eq.~\eqref{eq:S_int} is conserved by the RG procedure. 
\begin{figure}[tbh]
\includegraphics[width=1\columnwidth, clip]{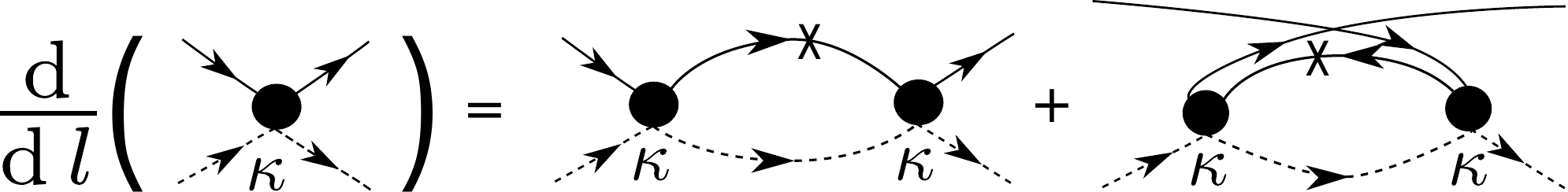}
\caption{\label{fig:rg_coupling_kappa} 
Graphical representation of Eq.~\eqref{eq:full_real_time}. 
The derivates with respect to $a$
of the electronic Green's functions are represented by crossed solid lines.}
\end{figure}
The renormalized coupling, $g'$ can thus be expressed as $g' = g + \delta g$, with 
$\delta g  = \delta g^{(1)}   + \delta g^{(1')}  $, and with $\delta g^{(1')} (t-t') $ given by an expression similar to 
\eqref{eq:deltag^(1)}. Introducing the scaling variable $l = \ln(a/a_0)$  and then dividing 
 $\delta g$ by $\delta a/a = \delta l$ we obtain an integro-differential equation for the coupling
 $g(t-t')$: 

\begin{widetext}

\bea
\frac{ {\mathrm d} g (t-t')}
{\mathrm{d}l}  &= &
\sum_{\tal,\ts,\tsi} \int {\rm dt_1\, dt_2}\,\left \{
g_{\al\tal}^{\s s;\tsi \ts}(t-t_1)g_{\tal\al'}^{\tsi \ts;\s' s'}(t_2-t')
\frac{\partial  G^{t}_{\tal}(t_1-t_2)}{\partial l}
F^{t}_{\ts}\bigl (\frac{t+t_1-t_2-t' }{2} \bigr)
e^{i {\lambda_s}(t_1-t')/2}
e^{-i {\lambda_{s'}}(t_2-t)/2}\right .
\nonumber
\\
&+&\left . g_{\al\tal}^{\s \ts;\tsi s'}(t-t_1)g_{\tal\al'}^{\tsi s;\s' \ts}(t_2-t')
\frac{\partial  G^{t}_{\tal}(t_1-t_2)}{\partial l}
F^{t}_{\ts}\bigl (\frac{t'+t_2-t_1-t }{2} \bigr)
e^{i {\lambda_s}(t_2-t)/2}
e^{-i {\lambda_{s'}}(t_1-t')/2} \right \}
\label{eq:full_real_time}
\eea
This constitutes a complete integro-differential equation for the vertex function with the 
boundary conditions \eqref{eq:g_initial}. It can be represented 
graphically as in Fig.~\ref{fig:rg_coupling_kappa}.
As we have discussed, during the RG procedure one generates pseudofermion self-energy corrections
(diagrams (d) in Fig.~\ref{fig:contractions}). The imaginary part of these self energy corrections corresponds to  pseudofermion decay, 
and is proportional to the spin relaxation rate  (see Sec.~\ref{sec:decoherence}). Eqs.~\eqref{eq:full_real_time}  thus incorporate 
spin relaxation through  the pseudofermion propagators, $F_s^t$. 

Fortunately, these somewhat cumbersome 
equations can be further simplified by 
relatively simple  approximations. 
At $T=0$ temperature, the pseudofermion propagator $F^t_s$ is approximately given by
\be 
F^{t}_{\ts} \Bigl (\frac{t+t_1-t_2-t' }{2} \Bigr)  \approx  - i \Theta \Bigl (\frac{t+t_1-t_2-t' }{2} \Bigr) e^{-i (t+t_1-t_2-t' )\lambda_\ts/2}. 
\ee
If we assume that   typical electronic time differences  involved in the vertices are short compared to 
$t_1-t_2$, we can  then set $t\to t_1$ and $t'\to t_2$ in the argument of the  $\Theta$ function,  giving $\Theta(t_1-t_2)$. 
With this approximation, the exponential functions can be regrouped and 
time integrals become simple convolutions. In this spirit, we approximate $F^{t}_{\ts} $ as
\be
F^{t}_{\ts} \Bigl (\frac{t+t_1-t_2-t' }{2} \Bigr) \approx \Bigl[F^{t}_{\ts} \bigl (t_1-t_2  \bigr) e^{i(t_1-t_2)\lambda_\ts }\Bigr] e^{-i (t+t_1-t_2-t' )\lambda_\ts/2}.
\label{eq:cut_off_function}
\ee
Writing furthermore  $G_\alpha^t(t)$ as $G_\alpha^t(t) = G_0^t(t) e^{-i\mu_\alpha t}$ and thus separating its trivial 
chemical potential dependence, 
the above integro-differential equations reduce 
to  the following differential equations in Fourier space, 
\begin{eqnarray}
\frac{ \rmd\; g_{\alpha\alpha'}^{ \sigma s; \sigma' s'} (\omega) }{\rmd l}  = 
& & \sum_{\tilde \alpha\tilde \sigma \tilde s}\left [ 
g_{\alpha \tilde \alpha}^{ \sigma \tilde s; \tilde \sigma s'} \left(\omega+
\frac{\lambda_{\tilde s s}}{2} \right) 
\;g_{\tilde \alpha \alpha'}^{ \tilde \sigma s; \sigma' \tilde s} \left(\omega+
{\lambda_{\tilde s s'} \over 2} \right)
\Xi_{\,a}^{\,\ts} 
\left(  \omega -  \frac{\lambda_{s \tilde s} + \lambda_{s' \tilde s}  }{2} 
 -\mu_{\tilde \alpha}     \right)
\right .\nonumber\\
& & \phantom{nnn} - 
\left . 
g_{\alpha \tilde \alpha}^{\sigma s ; \tilde \sigma \tilde s} \left(\omega+
\frac{\lambda_{s' \tilde s}}{2} \right) 
\;g_{\tilde \alpha \alpha'}^{\tilde \sigma \tilde s; \sigma' s'} \left(\omega+
\frac{\lambda_{s \tilde s}}{ 2} \right)
\Xi_{\,a}^{\,\ts} 
\left(  \omega +  \frac{\lambda_{s \tilde s} + \lambda_{s' \tilde s}  }{2} 
 -\mu_{\tilde \alpha}     \right)
\right ],
\label{eq:scaling_g}
\end{eqnarray}
\end{widetext} 
where the notation $\lambda_{ss'} = \lambda_s-\lambda_{s'}$ has been introduced for the energy splitting of the states $s$ and $s'$.
The cut-off function $\Xi_{\,a}^{\,s} 
  (\omega)$ can be expressed here as
\be
\Xi_{\,a}^{\,s}  (\omega)
= - \int \mathrm d t\, e^{i\omega t}
F^{t}_{s} ({t} ) \, e^{i t\lambda_s } 
\frac{\partial  G^{t}_0(t)}{\partial l}.
\label{eq:cut_off_function}
\ee
This function, on the one hand, accounts for the finite bandwidth of the conduction electrons, 
 and cuts off  contributions at frequencies $|\omega| \gtrsim 1/a$. However, it also accounts for 
 the finite temperature thermal decoherence of the conduction electrons at times $t>1/T$, and 
furthermore,  also incorporates the effect of  spin relaxation processes through 
 the pseudofermion propagator $F^t$.  For most practical purposes the detailed shape  of this cut-off 
 function  is not very relevant, and for  practical purposes it  can usually be replaced by a simple function 
\be
\Xi_a^s(\omega)\approx \Theta( 1 /a - (\omega^2+\Gamma^2)^{1/2}), 
\label{eq:cut_off_function}
\ee
with $\Gamma=\Gamma(a)$ a spin relaxation rate that we determine self-consistently
(for details, see Sec.~\ref{sec:decoherence}). ÊThe validity of these latter approximations can be checked against 
the solution of the full integro-differential equations, Eqs.~\eqref{eq:full_real_time}.
We emphasize that relaxation processes play an important role 
since even at $T=0$ temperature a finite bias  voltage can generate a large  intrinsic spin relaxation, 
$\Gamma$, which regularizes the logarithmic singularities.
The scaling equations  Eqs. \eqref{eq:scaling_g} are valid in the presence of the external magnetic field
which enters through the pseudofermions energy $\lambda_s = \lambda_0-sB/2$,  $(s=\pm1)$.
At the same time they are  identical to 
the equations  obtained in a more heuristic  way in Refs.~\onlinecite{Rosch.02} and~\onlinecite{Rosch.03}. However, in our 
real time functional RG formalism the derivation is rather straightforward and the
approximations  made are   better controlled.

Notice that  the usual poor man's RG
procedure can be  recovered by the local approximation, i.e. by dropping the  time-dependence of $g$, and replacing the generated 
non-local couplings 
by local ones,  ${\mathbf g}(t) \to  \delta(t)\; \int \rmd t\; 
{\mathbf g}(t)$, which corresponds to assuming frequency independent couplings 
in Eq.~\eqref{eq:scaling_g}. 

In the absence of an external magnetic field all couplings are of the form, 
\be 
g^{\sigma s;\sigma's'}_{\alpha\alpha'}(t)
= \frac 1 4 \;{\bs \sigma}_{ss'}\cdot \bs\sigma_{\sigma\sigma'}\;
g_{\alpha\alpha'} (t)\,,
\label{eq:full_couplings}
\ee
the terms $\lambda_{ss'}$ identically vanish, and and the renormalization group equations simplify to 
\begin{equation}
\frac{ \rmd\; g_{\alpha\alpha'} (\omega) }{\rmd l}  = 
 \sum_{\tilde \alpha}
g_{\alpha \tilde \alpha} \left(\omega\right) 
\;g_{\tilde \alpha \alpha'} \left(\omega\right)
\Xi_{\,a} 
\left(  \omega -\mu_{\tilde \alpha}     \right)
\label{eq:su2_scaling}
\end{equation}
\begin{figure}[tbh]
\includegraphics[width=1\columnwidth,clip]{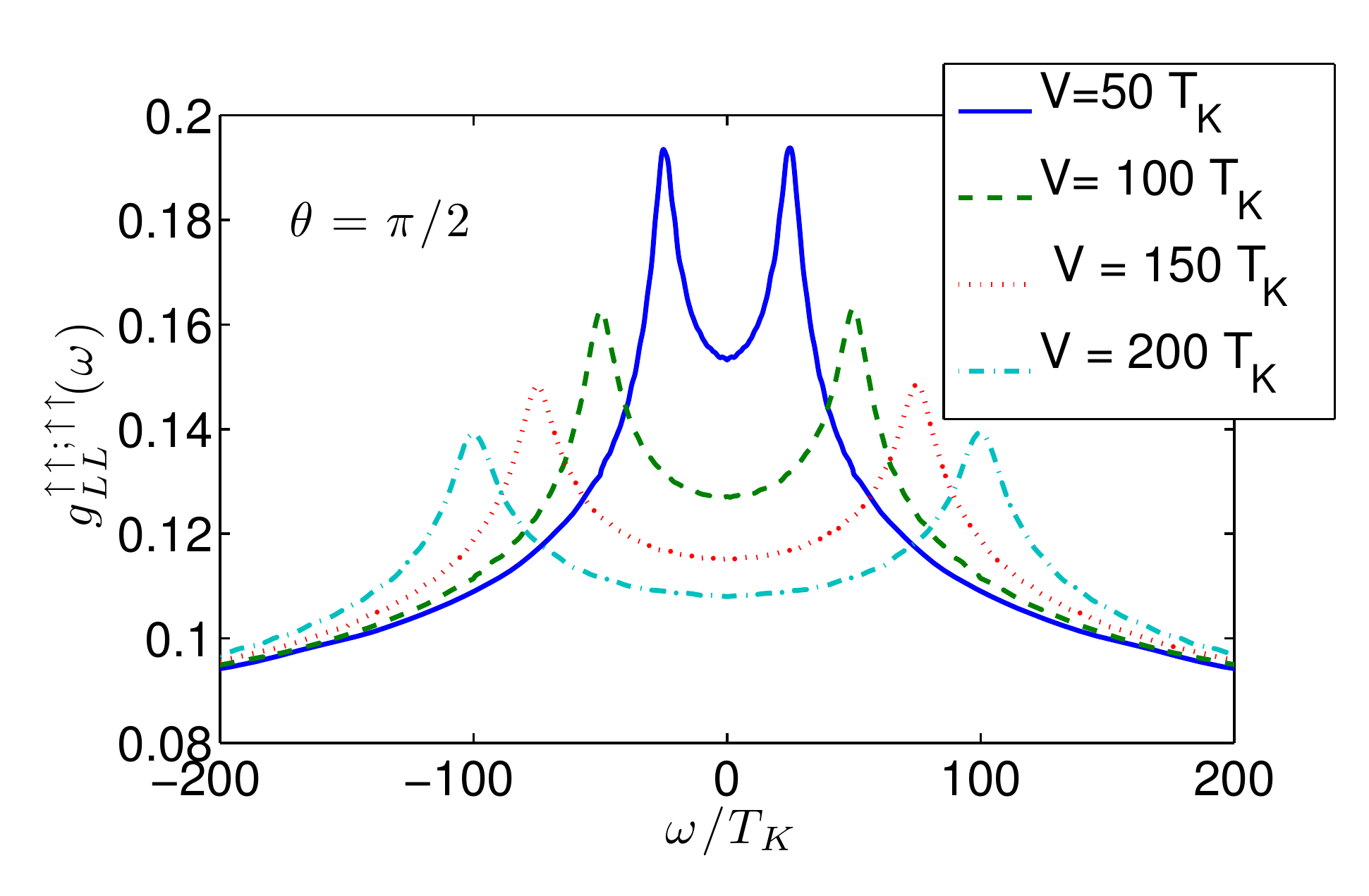}
\includegraphics[width=1\columnwidth,clip]{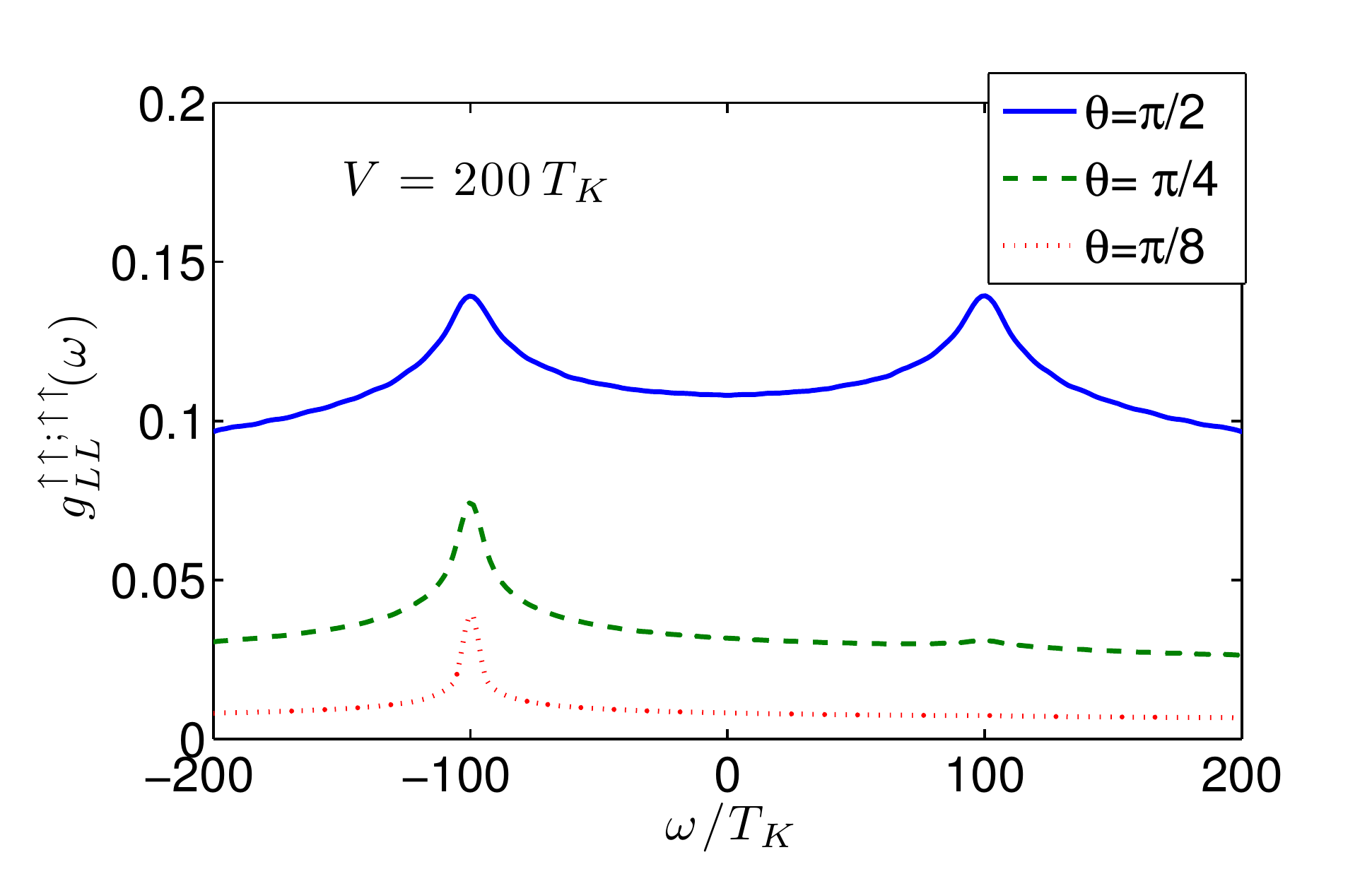}
\caption{\label{fig:couplings} (Color online)
Upper panel: 
The frequency dependence of the $g_{LL}^{\up\up;\up\up}(\w)$ component 
of the coupling matrix for different voltage biases. The magnetic field
is set to zero. Lower panel: The frequency dependence of the same component
for a fixed bias $V = 200\, T_K$, but for different $\theta$. Spin relaxation has been incorporated self
consistently (see text). 
}
\end{figure}
\begin{figure}[tbh]
\includegraphics[width=1\columnwidth,clip]{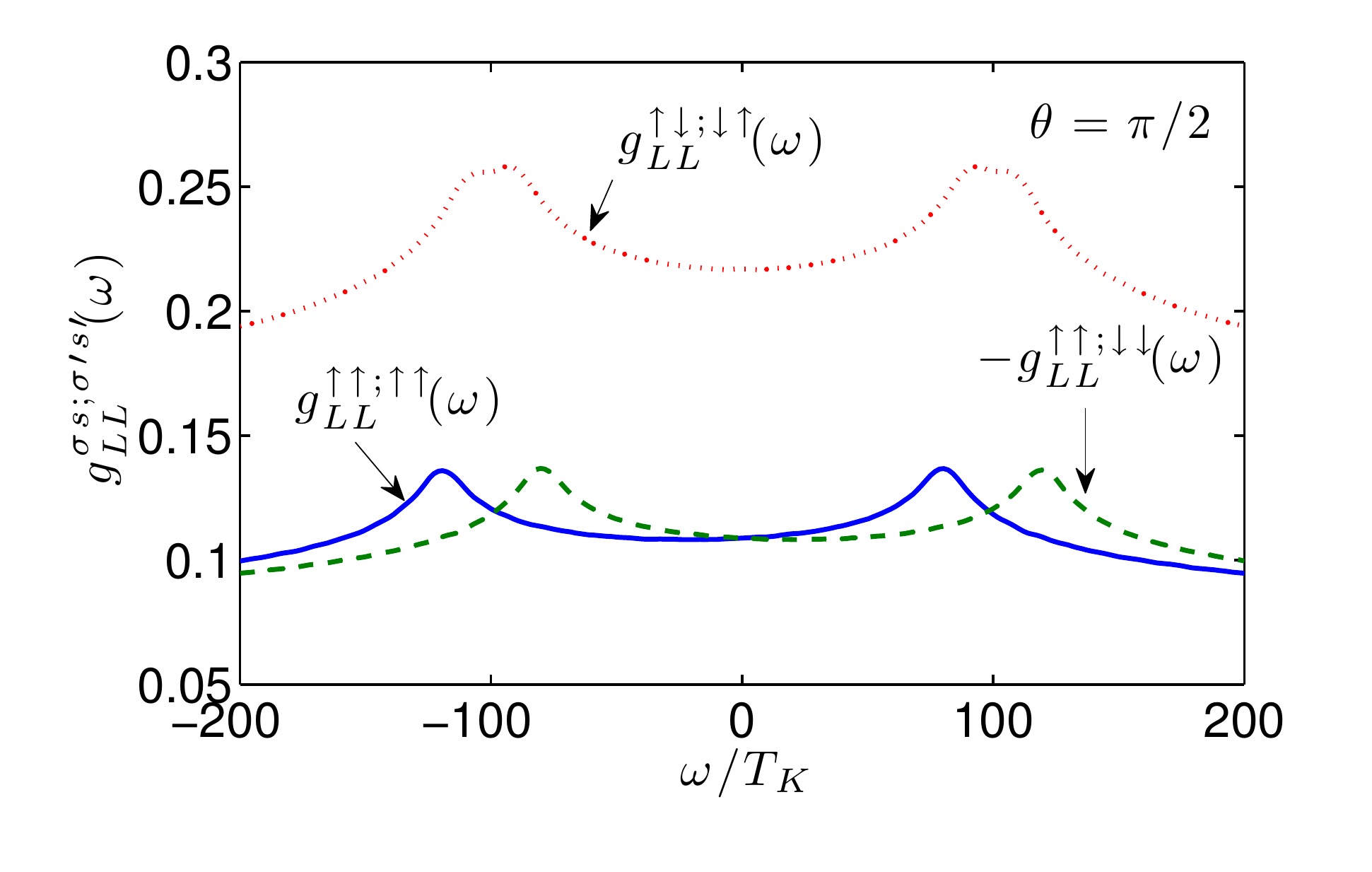}
\caption{\label{fig:couplings_B} (Color online)
The frequency dependence of the non-zero components of the coupling matrix 
in an external magnetic field $B = 20\,T_K$. The voltage bias is
fixed to $V = 200\, T_K$.}
\end{figure}

We determine the renormalized couplings by solving  Eq. \eqref{eq:scaling_g} 
[or Eq.~\eqref{eq:su2_scaling}]
numerically,  while  taking into account the spin decoherence rate (see Sec.~\ref{sec:decoherence} and particularly  Eq. \eqref{eq:decoherence}). 
As discussed in the introduction, the initial couplings $j_{\alpha\beta}$ can be  parametrized in terms of a single dimensionless 
coupling, $j$, and a spinor $v_\alpha$. It is convenient to choose a suitable gauge so that  $v_\alpha$ is real, and can be parametrized 
in terms of a single  angle $\theta$, as $\{v_\alpha\} = \{\cos(\theta/2),\sin(\theta/2)\} $. In terms of this, the matrix  $j_{\alpha\beta}$ becomes
\be 
{\bf j} = 
\begin{pmatrix}
j_{LL} & j_{LR}  \\
j_{RL} & j_{RR}
\end{pmatrix}
=
j
\begin{pmatrix}
\cos^2 {\theta \over 2} &   \cos{\theta \over 2}\sin {\theta \over 2}  \\
 \cos {\theta \over 2} \sin {\theta \over 2} & \sin^2 {\theta \over 2}
\end{pmatrix}
\;.
\label{eq:j_theta}
\ee
The value  $\theta=\pi/2$ corresponds to symmetrical coupling to the left and right electrodes, while 
for $\theta = \pi$  and $\theta=0$ the quantum dot is decoupled from the left  and right electrode, respectively.

Typical results for some of the components of the matrix ${\mathbf g} (\omega)$ are 
displayed in Fig.~\ref{fig:couplings}. The 
rescaled couplings display strong features (logarithmic singularities in the 
absence of decoherence) at
frequencies of the order of voltage drop between the external contacts, 
$\omega =\pm V/2$. 
The effect of the asymmetry on the renormalized couplings is shown  in Fig.
\ref{fig:couplings}. 
Notice that for strong   asymmetry ($\theta\approx 0$ or $\theta\approx \pi$),
Êonly the peak associated with the Fermi surface of the  more strongly coupled electrode 
survives,  while the other  is almost washed away. The effect of the magnetic field on 
the renormalized couplings is presented in Fig.~\ref{fig:couplings_B}. The 
longitudinal couplings $g_{\alpha\beta}^{\s s;\s s }(\w)$ 
develop peaks at $\omega = \pm V/2-\sigma B$, while the transversal ones $g_{\alpha\beta}^{\s \bar \s;\bar \s \s }(\w)$
develop peaks at  at $\pm 1/2(V\pm B)$. Notice that 
the decoherence rate $\Gamma$ 
 prevents the flow from running towards the 
strong coupling regime. Therefore, the peaks in the renormalized couplings get slightly
broader and partially suppressed.

\section{Scaling equations for the current vertex operator}
\label{sec:current_scaling}

\subsection{Current and noise definitions}\label{sec:noise}\label{sec:current_noise}

Having established the RG equations for the interaction vertex, let us now turn to the definition and 
renormalization of the current operator. The current operator can be constructed by exploiting the equations of motion, and  take on forms similar to  Eqs.~\eqref{eq:H_int} and \eqref{eq:H_int_gen}.  In the Kondo model one trivially finds 
\be
\hat I_L (t) = - \hat I_R (t)  = \sum_{\alpha\beta}\frac e 
2 v^L_{\alpha\beta} \, {\mathbf S}(t) \cdot
 \psi^\dagger_{\alpha}(t){\bs\sigma} \psi_{\beta}(t)\, ,
\label{eq:current_operator}
\ee
with the time arguments indicating Heisenberg operators. Here, for simplicity, we suppressed  the internal spin indices  and expressed 
the current vertex matrices as 
\be 
{\bf v}^L \equiv {\bf v} = -{\bf v}^R=
\begin{pmatrix}
0 &  - i j_{LR}  \\
 i j_{LR} & 0
\end{pmatrix}
\;.
\ee
In the Kondo model current conservation is satisfied  at the operator level, $\hat I_L (t) = - \hat I_R (t)$.
For the general Hamiltonian, \eqref{eq:H_int_gen}, the equation of motion amounts to a similar expression of the form,
\be
\hat I_k (t)   = \sum_{m,n, s,s'} e \,
 [v^{k}]_{mn}^{ss'} X_{ss'}(t) 
 \psi^\dagger_{m} \psi_{n}(t)\, ,
\label{eq:current_operator}
\ee
with the $X_{ss'}(t) $ denoting time evolved Hubbard operators, $|s\rangle\langle s'| $, and the current vertices given by 
\be
 [v^{(k)}]_{mn}^{ss'} = - i \, (\delta_{mk} -  \delta_{nk}) g_{mn}^{ss'}\,.
 \ee

Having the current operators at hand, one can then define the various  current-current correlation functions. 
The 'bigger' and 'lesser' noise correlation functions are defined as
\bea
S^>_{\alpha\beta} (t,t') &=&  \langle \hat I_\alpha (t)\,   \hat I_\beta (t')\rangle \mbox{ and }
\label{eq:up-noise}
\\
S^<_{\alpha\beta} (t,t') & = &  \langle \hat I_\alpha (t')\,  \hat I_\beta (t)\rangle, 
\label{eq:low-noise}
\eea
while  the symmetrized and antisymmetrized noise components are given by
\begin{equation}
S^{s/a}_{\alpha\beta} (t,t') =  \frac{1}{2} \langle [ \hat I_\alpha (t)\, ,  \hat I_\beta (t')]_{\pm}\rangle,
\label{eq:sym-noise}
\end{equation}
with $[.., ..]_\pm$ denoting  anticommutators/commutators, respectively.
The Fourier spectra of these are directly accessible through noise measurements.  
The spectra of $S^{\gtrless}(\omega)$ can be measured by emission or absorption 
experiments,~\cite{Billangeon.06, Aguado.00} 
while the symmetrized noise is accessible through  standard a.c. noise 
spectroscopy.~\cite{ Gavish.00, Reulet.08}

\subsection{Current vertex scaling equations}\label{sec:current_vertex}

Having established the RG equations for the interaction vertex, let us now turn to the 
renormalization of the current operator. To compute the 
current-current correlation functions \eqref{eq:low-noise} and
\eqref{eq:sym-noise} within the path integral formalism,   we 
first  express the current operators $\hat I_{\alpha}$ in terms of Grassmann fields
on the Keldysh contour, $\hat I_{\alpha}\to I^{(\kappa)}_{\alpha}$. Representing the spin operators using 
pseudofermions we obtain 
 \bea
 \hat I_{\alpha}\to I^{(\kappa)}_{\alpha} (t) &=& 
 \sum_{\alpha\alpha' \sigma\sigma' s s'} \frac e 
4 v^\alpha_{\alpha\alpha'}  \bar f^{(\kappa)}_s(t) {\bs \sigma}_{ss'} f^{(\kappa)}_{s'} (t) 
\nonumber
\\
&\cdot &
 \bar \psi^{(\kappa)}_{\alpha\sigma }(t){\bs\sigma}_{\sigma\sigma'} \psi^{(\kappa)}_{\alpha'\sigma'}(t)\, .
 \label{bare_current}
\eea
Introducing then  the corresponding generating functional, 
\be 
Z[h^{(\kappa)}_{\alpha}(t)] \equiv 
\langle e^{-i \;\sum_{\alpha=L,R} \sum_\kappa \int \rmd t \;  h^{(\kappa)}_\alpha(t) I^{(\kappa)}_{\alpha} (t)}\rangle_{\cal S}\;,
\label{generating_functional}
\ee
all current-current correlation functions can be generated by functional differentiation with respect to $h_\alpha(t)$.

Unfortunately, however, similar to the interaction vertex ${\bs g}$, the current vertex becomes non-local in time upon rescaling 
$a\to a'$, and therefore Eq.~\eqref{bare_current} only holds for the {unrenormalized} (bare) current operator. 
Also,  though the current vertex is initially obviously related to the interaction vertex, 
its generated time structure turns out to be very different from that of the interaction vertex and,  
as we show  later, it necessarily acquires the following form under the RG
\bea 
I_L(t) &=& e \sum_{\alpha\alpha'}\sum_{ \sigma \sigma'ss'}
\int {\rm d}t_1\,{\rm d}t_2\;(V^L)_{\alpha\alpha'}^{ \sigma s; \sigma' s'}(t_1-t,t-t_2)\;
\nonumber
\\
&& \bar f_{s}(t) f_{s'}(t)  \cdot
\bar \psi_{\alpha\sigma}(t_1)  \psi_{\alpha'\sigma'}(t_2) \;
\label{eq:nonlocal_current}
\eea
with the initial condition that for the bare theory, $a=a_0$
\be
(V^L) _{\alpha\alpha'}^{ \sigma s; \sigma' s'}(\tau_1,\tau_2,a_0) = \frac 1 2 \delta(\tau_1)\;\delta(\tau_2)
\;{ v}^L_{\alpha\beta}  \, {\bs \sigma}_{\s \s'} {\mathbf S}_{ss'}.
\label{eq:vertex_initial}
\ee
The structure \eqref{eq:nonlocal_current} follows from  detailed derivations, however, heuristically
 one can argue that this structure is needed to 
keep track of the time of \emph{measurement $t$} in addition to the times  where the incoming and outgoing electrons are scattered,
$t_1$ and $t_2$ (see the Feynman diagrams in Fig.~\ref{fig:diagram_components}).  
Although the current vertex still has a four-leg structure, just as the vertex, 
due to  its double time dependence 
it can no longer be identified with the coupling vertex. Therefore, we represent it by a different diagrammatic symbol,
 depicted in Fig. \ref{fig:diagram_components}.

\begin{figure}[h]
\includegraphics[width=1\columnwidth,clip]{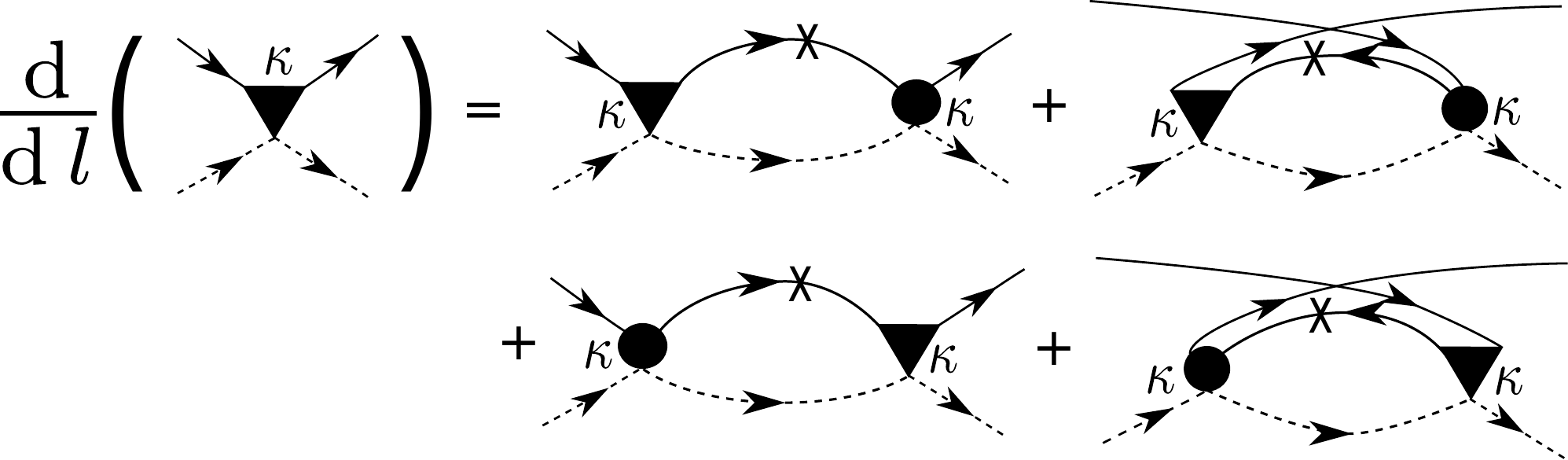}
\caption{\label{fig:rg_current_kappa} 
Diagrams renormalizing  the current vertex. 
Crosses indicate the logarithmic  derivative with respect to the 
scaling parameter $a$. Notice that the current and 
the coupling vertex do not "commute" and the diagrams came in pairs. 
Each diagram corresponds to a term in Eqs. \eqref{eq:full_V_real_time}
and \eqref{eq:scaling_current}.}
\end{figure}

To investigate how the current vertex is renormalized, we follow a strategy similar to that of Section~\ref{sec:scaling_g}. 
We expand the generating functional \eqref{generating_functional} simultaneously in the field $h_{\alpha}^{(\kappa)}(t)$ and 
also in the interaction kernel, $\mathbf{g}(t)$. Again, similar to Sec.~\ref{sec:real_time_approach} we find that changing $a\to a'$
changes only the contributions of those diagrams and those configurations, where (at least) two contracted fields are close to each 
other, $|t_1-t_2| \sim a$.  The contractions of the electronic fields appearing  in a current vertex can then be classified similar to Fig.~\ref{fig:contractions}, and  one can argue that only contractions shown  in Fig.~\ref{fig:rg_current_kappa} must be considered, with both the current and the interaction vertex lying  on the same branch of the Keldysh contour.  
Notice that the current and the coupling vertex are not equivalent 
and therefore do not "commute" when  placed on one of the branches of the 
Keldysh contour. Therefore the number of diagrams for the upper  branch of the Keldysh contour is four.
These diagrams are sketched in Fig. \ref{fig:rg_current_kappa}.

The RG equations of the current vertex can be obtained following very similar lines as in Sec.~\ref{sec:real_time_approach}, and 
we obtain for the left current vertex, $V_{\alpha\alpha'}^{\sigma s; \sigma' s'}\equiv (V^L)_{\alpha\alpha'}^{\sigma s; \sigma' s'}$
\begin{widetext}
\bea
\frac{ {\mathrm d} V_{\alpha\alpha'}^{\sigma s; \sigma' s'} (t-\tilde t,\tilde t- t')}
{\mathrm{d}l}  &= &
\sum_{\tal,\ts,\tsi} \int {\rm dt_1\, dt_2}\,\left \{
g_{\al\tal}^{\s s;\tsi \ts}(t-t_1)V_{\tal\al'}^{\tsi \ts;\s' s'}(t_2-\tilde t, \tilde t - t' )
\frac{\partial  G^{t}_{\tal}(t_1-t_2)}{\partial l}
F^{t}_{\ts}\bigl (\frac{t+t_1 }{2} -\tilde t \bigr)
e^{i {\lambda_s}(t+t_1- 2\,\tilde t)/2}\right .
\nonumber
\\
&+&\left . g_{\al\tal}^{\s \ts;\tsi s'}(t-t_1)V_{\tal\al'}^{\tsi s;\s' \ts}(t_2-\tilde t, \tilde t-t')
\frac{\partial  G^{t}_{\tal}(t_1-t_2)}{\partial l}
F^{t}_{\ts}\bigl (\tilde t-\frac{ t_1+t}{2} \bigr)
e^{-i {\lambda_{s'}}(t_1+t-2\, \tilde t)/2} \right.\nonumber
\\
&+&\left . V_{\al\tal}^{\s \ts;\tsi s'}(t-\tilde t, \tilde t - t_1 )
g_{\tal\al'}^{\tsi s;\s' \ts}(t_2-t')
\frac{\partial  G^{t}_{\tal}(t_1-t_2)}{\partial l}
F^{t}_{\ts}\bigl (\frac{t'+t_2 }{2} -\tilde t \bigr)
e^{i {\lambda_s}(t_2+t'- 2\,\tilde t)/2}\right .\nonumber
\\
&+&\left . V_{\al\tal}^{\s s;\tsi \ts}(t-\tilde t, \tilde t - t_1 )
 g_{\tal\al'}^{\tsi \ts;\s' s'}(t_2-t')
\frac{\partial  G^{t}_{\tal}(t_1-t_2)}{\partial l}
F^{t}_{\ts}\bigl (\tilde t -\frac{t'+t_2 }{2}  \bigr)
e^{-i {\lambda_{s'}}(t_2+t'- 2\,\tilde t)/2}\right \}
\label{eq:full_V_real_time}
\eea

Again, following the same steps as in Sec.~\ref{sec:scaling_g}
and approximating the cut-off function as in \eqref{eq:cut_off_function}
we obtain, 
\begin{eqnarray}
\frac{ \rmd\; V_{\alpha\alpha'}^{\sigma s;\sigma' s'} (\omega, \omega') }{\rmd l} & = &
- \sum_{\tilde \alpha\tilde \sigma \tilde s}\Bigl [
g_{\alpha \tilde \alpha}^{\sigma s; \tilde \sigma \tilde s} \left(\omega+
{\lambda_{s\ts } \over 2} \right) 
\;V_{\tilde \alpha \alpha'}^{ \tilde \sigma \tilde s;  \sigma' s'} \left(\omega+
\lambda_{s\ts} , \omega' \right)
\Xi^{\tilde s}_a 
\left( \omega +   \lambda_{s\ts }   -\mu_{\tilde \alpha}     \right)
\nonumber\\
& & + 
 V_{\alpha \tilde \alpha}^{\sigma s; \tilde \sigma \tilde s} \left(\omega, 
\omega'+\lambda_{s'\ts }\right) 
\;g_{\tilde \alpha \alpha'}^{ \tilde \sigma \tilde s;  \sigma' s'} \left(\omega'+
{\lambda_{ s'\ts} \over 2}  \right)
\Xi_a^{\ts} 
\left(\omega' +\lambda_{s'\ts} 
 -\mu_{\tilde \alpha}  \right )  \nonumber\\
& & -
g_{\alpha \tilde \alpha}^{ \sigma \tilde s; \tilde \sigma s'} \left(\omega+
{\lambda_{\tilde ss'} \over 2} \right) 
\;V_{\tilde \alpha \alpha'}^{ \tilde \sigma s;  \sigma' \tilde s} \left(\omega+
\lambda_{\tilde ss'}, \omega' \right)
\Xi_a^\ts
\left(\omega +   \lambda_{\tilde s s'}   -\mu_{\tilde \alpha}       \right)
\nonumber\\
& & -
 V_{\alpha \tilde \alpha}^{ \sigma \tilde s; \tilde \sigma s'} \left(\omega, 
\omega'+\lambda_{\tilde s s}\right) 
\;g_{\tilde \alpha \alpha'}^{ \tilde \sigma s;  \sigma' \tilde s} \left(\omega'+
{\lambda_{\tilde s s} \over 2}  \right)
\Xi_a^\ts
\left(\omega' +\lambda_{\tilde s s} 
 -\mu_{\tilde \alpha}     \right)
\Bigr ].
\label{eq:scaling_current}
\end{eqnarray}
\end{widetext}
This set of  equations needs be solved  parallel to the scaling equations,
Eq.~\eqref{eq:scaling_g} with the initial condition
\eqref{eq:vertex_initial}.

\begin{figure}[htb]
\includegraphics[width=1\columnwidth,clip]{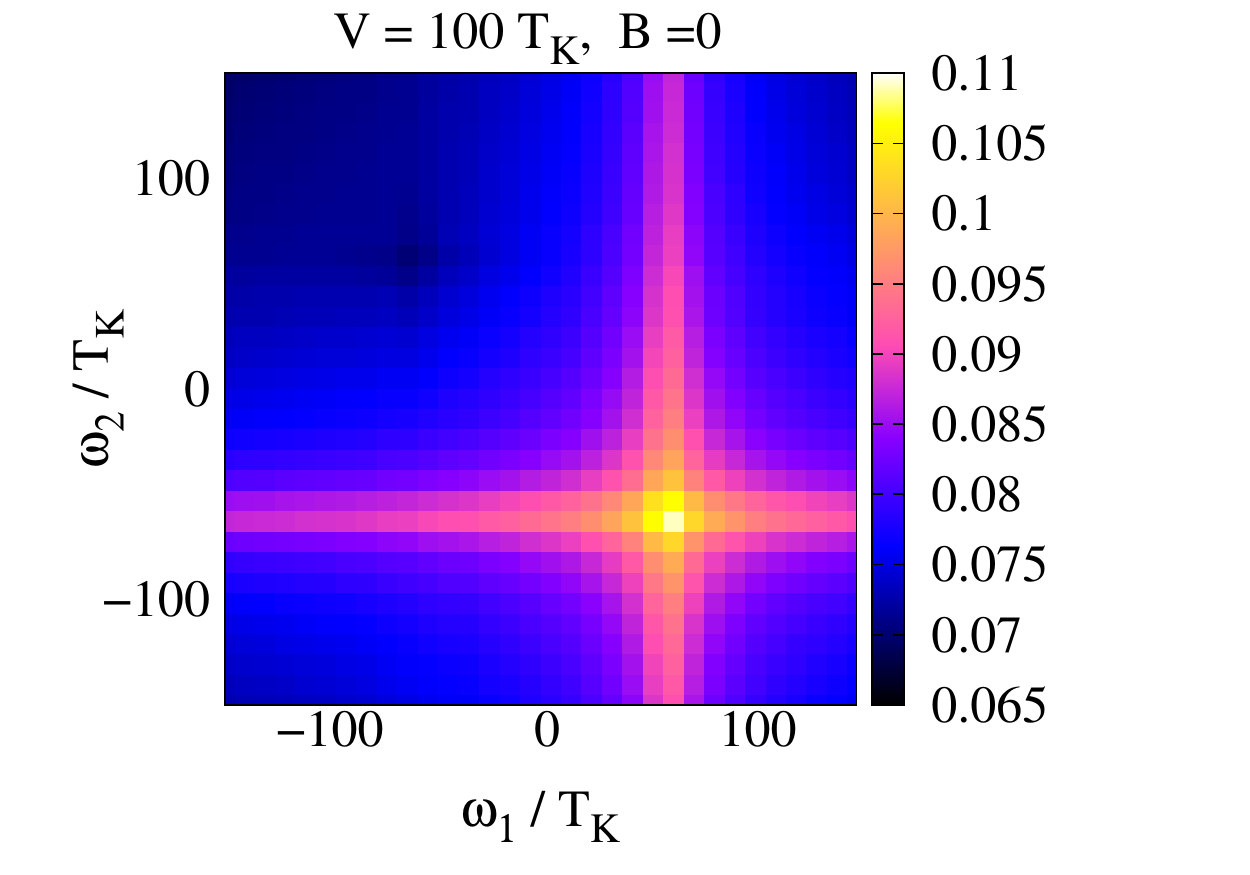}
\includegraphics[width=1\columnwidth,clip]{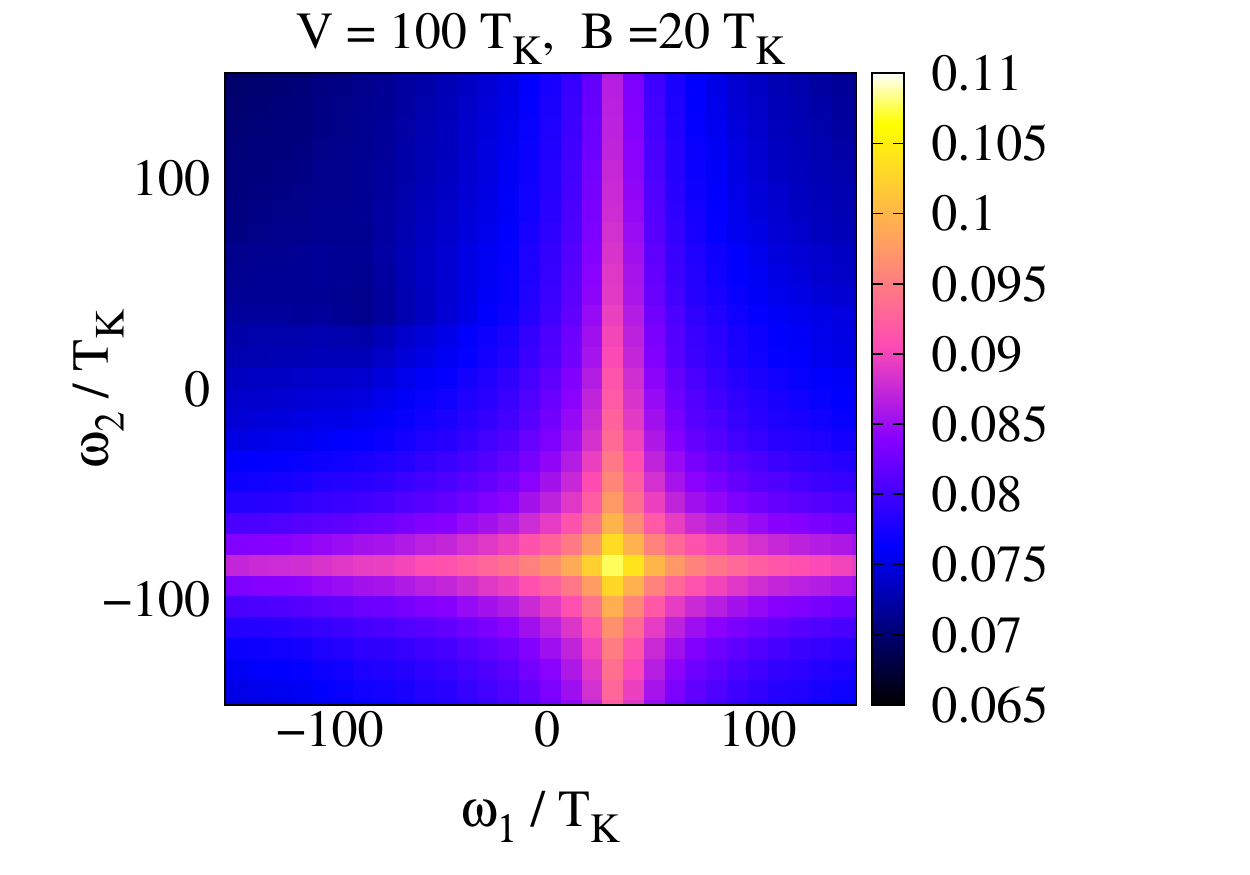}
\caption{ (color online) Frequency dependence of the current vertex kernel 
$V_{LR}^{ \uparrow\uparrow; \uparrow\uparrow} (\omega_1, \omega_2)$  
for a fixed voltage bias $V=50\; T_K$, and 
for magnetic field B =0 (upper panel) and $B=20\; T_K$ (lower panel). In the 
absence of an external field it shows a logarithmic singularity at frequencies $
\hbar\omega\simeq \pm eV/2$, which is then shifted by the presence of the 
external magnetic field. } 
\label{fig:current_vertex}
\end{figure}


As we discussed before,  though 
the renormalized couplings ${\bf g}(\omega)$ drive the scaling 
of the current vertices, ${\bf V} (\omega_1,\omega_2,a_0)$, there seems
 to be no simple connection between these two. In other words, 
it is unavoidable to introduce the renormalized current vertices within 
the  functional RG scheme to compute time-dependent 
current correlations. Very importantly, the above extension  also solves to problem of  {\it current conservation}:  
Eq.~\eqref{eq:scaling_current} is   linear in $ {\bf V}$. Therefore, as the bare vertices satisfy
$\mathbf{v}_L = -\mathbf{v}_R$, the full vertices shall also satisfy
$\mathbf{V}_L(t,t') = -\mathbf{V}_R(t,t')$ by construction, and therefore the condition 
$I^{(\kappa)}_L(t) + I^{(\kappa)}_R(t)\equiv 0$ is automatically fulfilled 
for any value of the cut-off, $a$.  On the other hand, 
we could not find any systematic way to generate a current field from just the renormalized action, Eq.\eqref{eq:S_int}, such that it respected 
current conservation.  The introduction of the current vertex and its RG equation   
seem to be  therefore unavoidable to extend the formalism of Refs.~[\onlinecite{Rosch.03, Rosch.01}] to compute time dependent correlations.

Fig.~\ref{fig:current_vertex} displays 
 the frequency dependence of the  $\uparrow\uparrow;\uparrow\uparrow$ component for a fixed bias voltage,
 in the absence/presence of the external magnetic field. 
Similar to the renormalized couplings, the components 
of the current vertex display  logarithmic singularities in the
frequency space at $\omega=\pm V/2$. These singularities  get shifted 
in the presence of the external magnetic field. 

\section{Decoherence effects}
\label{sec:decoherence}

Let us now turn to the important issue of decoherence.  Under non-equilibrium circumstances, 
 a large bias voltage necessarily entails a finite spin lifetime and related decoherence effects,  as also  observed
 experimentally.~\cite{Paaske.06,Basset.12} These decoherence effects  lead to a natural low energy cut-off for the logarithmic 
 singularities and the renormalization group flow. 
 In this section, we try to capture spin relaxation within two different approaches:  First  we  use a perturbative 
 master equation method  with renormalized couplings to determine the voltage and temperature dependence of the Korringa 
 spin relaxation rate, $\Gamma_K(T,V)$.   Then, in subsection~\ref{sec:pseudofermion_relaxation} we compute the pseudofermion's relaxation rate, $\Gamma_{\rm pf} (V,T)$. Both approaches result in a consistent picture when combined self-consistently
with the RG scheme developed in Sections~\ref{sec:real_time_approach} and \ref{sec:current_scaling}.

\subsubsection{Korringa relaxation rate: a Master equation approach} 
In the perturbative regime, $\min\{T, eV, B\}\gg T_K$, 
 one can investigate the relaxation of the  spin  by   perturbation theory.
In this parameter range, spin flip events are rare, and they  can be treated as a 
Markov process.~\cite{Pustilnik.05} 
The scattering events in this Markov process are generated by the exchange interaction, and  
consist of the scattering of an electron with spin $\s'$  from lead $\beta$ into 
a final state  of spin $\s$ in lead $\alpha$, while flipping the impurity spin 
 from $s'$ to $s$.  To leading order in perturbation theory, the transition rate for such process 
  $\gamma_{\alpha\leftarrow\beta}^{\s s\leftarrow \s' s'}$, is given by the simple Fermi Golden rule expression, 
\begin{multline}
\gamma_{\alpha\leftarrow\beta}^{\s s\leftarrow \s' s'} \approx   \frac \pi 2 \int \rmd \omega\,
\mid  j_{\alpha\beta}\,{\boldsymbol \s}_{\s\s'}\,{\mathbf S}_{ss'}  \mid^2
\bar f_{\alpha}(\omega + \lambda_{ss'})
 f_{\beta}(\omega)
 \label{eq:decoherence}
\end{multline}
with $f_{\alpha}(\omega) = f(\omega-\mu_{\alpha}) $ the shifted   Fermi-Dirac distribution for the electrons in the lead $\alpha$, and 
$\bar f_\alpha (\omega) = 1-f_\alpha(\omega)$. Notice the shift in the energy of the conduction electrons by $ \lambda_{ss'} = -(s-s') B/2$ in the argument 
due to energy conservation.  Within this simple master equation approach it then follows that 
the  spin  decays exponentially, 
\be
\langle S_z \rangle \propto \exp\,\{-(\Gamma_{\Uparrow} + \Gamma_{\Downarrow})t\} \ee
with:
\be\Gamma_{\Uparrow} = \sum_{\alpha, \beta,\s,\s'} \gamma_{\alpha\leftarrow\beta}^{\s\Uparrow
\leftarrow \s'\Downarrow}\;\;
{\rm and}\;\; \Gamma_{\Downarrow} = \sum_{\alpha, \beta,\s,\s'} \gamma_{\alpha\leftarrow\beta}^{\s\Downarrow
\leftarrow \s'\Uparrow}.\ee
This allows us to identify the 
Korringa relaxation rate  
\be
\Gamma_K (T, eV, B) =\Gamma_{\Uparrow}+\Gamma_{\Downarrow},
\label{eq:Korringa}
\ee 
as the relevant decoherence rate in the problem.

The integrals \eqref{eq:decoherence} can be evaluated analytically. In the limit of large voltages, $eV\gg T, B$, e.g., 
$\Gamma_K$ assumes  a simple analytical form:
\begin{multline}
\Gamma_K(T, eV\gg B) \approx  \pi T \left(
j_{LL}^2+j_{RR}^2 \right ) +\\  
j_{LR}^2\, eV\, \coth\left( {eV\over 2 T}\right)
+ {\cal O}(B^2)\;.
\end{multline}

For large magnetic fields, on the other hand, $B\gg eV$,  we obtain
\begin{multline}
\Gamma_K(T, eV\ll B) = \pi {B \over 2} \left(
j_{LL}^2+j_{RR}^2 + 2j_{LR}^2 \right )
\coth\left( {B\over 2 T}\right)\\ 
+ {\cal O}(eV^2)\;.
\end{multline}

It is instructive to express the Korringa rate in terms of the anisotropy angle, $\theta$, introduced in Eq. \eqref{eq:j_theta}.
In the limit when one of the variables $T$, $B$, or $V$ is much larger than 
the other two we obtain: 
\be 
\Gamma_K =
\begin{cases}  \pi j^2 T   & \mbox{\phantom{nnn}} (B,V\to 0) \;,\\
{1\over 4}\pi j^2 |eV|\sin^2(\theta)   &\mbox{\phantom{nnn}} (T,B\to 0)\;, \\
{1\over 2}\pi j^2 |B| &\mbox{\phantom{nnn} } (T,V \to 0) \; .
\end{cases} 
\label{eq:gammaK}
\ee
%
These results agree with those of Refs.~\onlinecite{Paaske.04,Kehrein.05a,Schoeller.09}. Notice that  the temperature and the magnetic field generate a 
decoherence rate  independent of the asymmetry, while  
the voltage-induced spin relaxation rate depends on the asymmetry. This is not so surprising
since the current flowing through the device is proportional to $\sim \sin^2(\theta)$, 
and is suppressed for a strongly asymmetrical quantum dot.  The Korringa rate in the large voltage limit is directly 
proportional to this current and thus strongly depends on the asymmetry $\theta$.

The previous results were obtained to lowest order in the exchange coupling, $j_{\alpha\beta}$. 
Higher order logarithmic corrections can be summed up perturbatively.~\cite{Paaske.04,Schoeller.09}
We can estimate the size of these corrections  by simply replacing the bare coupling $j$ with
 its renormalized value, $j\rightarrow 1/\ln(\max(|eV|, |B|, T)/T_K)$.  This approximation, however, breaks down 
 at $V\approx 10\, T_K$. In order to approach the regime $V\approx T_K$,
the self-consistent incorporation of the relaxation rate $\Gamma_K(V)$ is necessary.~\cite{Pletyukhov.12}
In particular, in the FRG scheme we follow Ref.~\onlinecite{Paaske.04} to 
express the running value of the relaxation rate $\Gamma_K(a)$ as 
\begin{multline}
\Gamma_K(a) ={2\pi} \sum_{\alpha\beta\s\s'}\sum_{s\ne s'}
\int{\rm d}\omega\, |g_{\alpha\beta}^{\s s; \s's'}(a, \omega+\frac{\lambda_{ss'}}{2})|^2\times\\
\bar f_{\alpha}(\omega+\lambda_{ss'}) f_\beta(\omega)
\end{multline}
The physical Korringa rate $\Gamma_K$ is then obtained by 
solving the RG equation self-consistently with the cut-off function, 
\eqref{eq:cut_off_function} and taking the $a\to 0$ limit.

\begin{figure}[h]
\includegraphics[width=\columnwidth,clip]{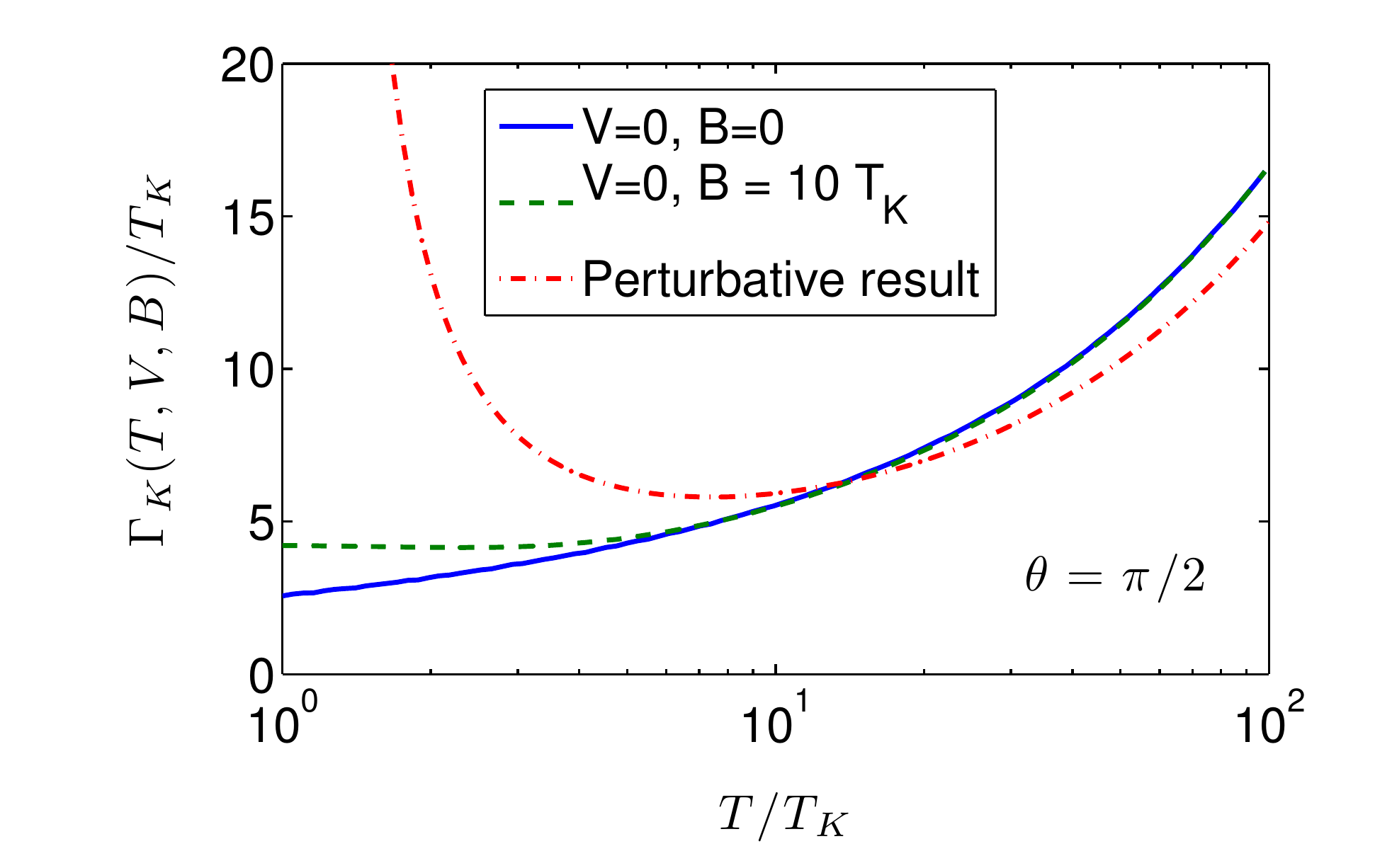}
\caption{(Color online)
The temperature dependence of the Korringa rate for 
V = 0. The dashed-dotted (red) line
represents the perturbative result: $\Gamma_K(T\gg eV, B) \simeq \pi\, T /\ln^2(T/T_K)$.
\label{fig:Korringa_rate_T}
}
\end{figure}

In Figs.~\ref{fig:Korringa_rate_V} and \ref{fig:Korringa_rate_T} 
we compare the voltage and temperature 
dependence of the Korringa rates as computed perturbatively and by the FRG method.
Clearly, for  $V\gg T_K$
the perturbative result gives a good estimate, 
but the result starts to deviate below $V \simeq 10\, T_K$, and  for $V\approx T_K$ a self-consistent calculation
of $\Gamma_K$ is necessary.  The physical explanation of the much better performance of FRG is simple: 
approaching $T_K$ the effective exchange rate becomes large. This, however, generates an increased spin relaxation rate, 
which then naturally feeds back and provides a cut-off for the logarithmic divergency. 

Notice that our real time FRG results are restricted to the region 
where at least one of the parameters $T$, $eV$, or $|B|$ is somewhat larger than $T_K$. Therefore,  
the $B=T=0$ FRG curve in Fig.~\ref{fig:Korringa_rate_V} (the $V=B=0$ FRG curve in 
Fig.~\ref{fig:Korringa_rate_T})  should be considered wit a grain of 
salt for voltages $|eV|\sim T_K$ (temperatures $T\sim T_K$).

\begin{figure}[h]
\includegraphics[width=1.0\columnwidth,clip]
{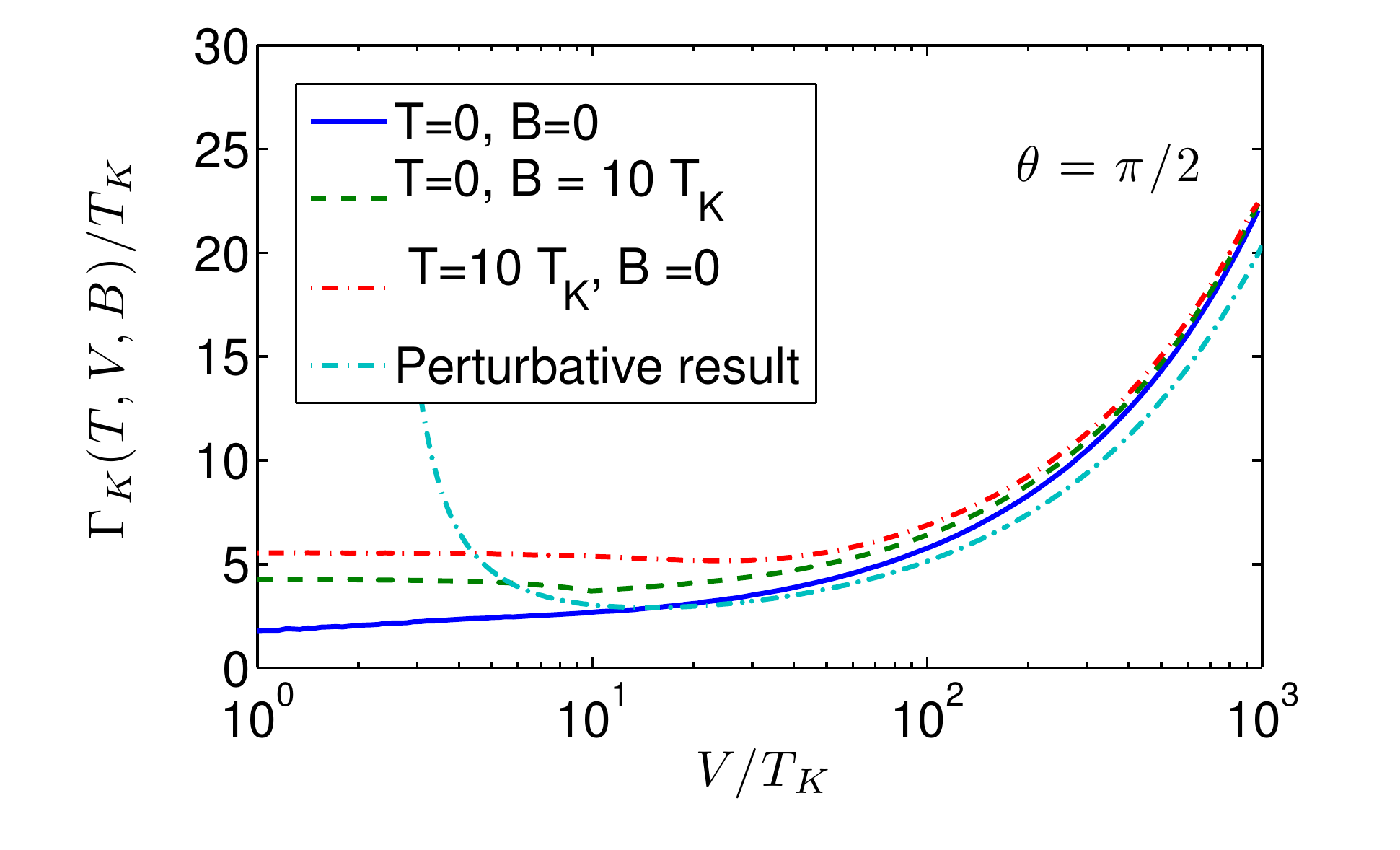}
\caption{\label{fig:Korringa_rate_V} (Color online)
The voltage dependence of Korringa rate for 
different temperatures and magnetic fields. The dashed (blue) line
represents the perturbative result: $\Gamma_{K}(eV\gg B, T) = \pi |eV|/4 \ln^2(|eV|/2\,T_K)$.
\label{fig:Korringa_rate_V}
}
\end{figure}

\subsubsection{Pseudofermion self-energy and lifetime}
\label{sec:pseudofermion_relaxation}

Within  the FRG scheme, 
one often identifies the pseudofermion relaxation rate $\Gamma_{\rm pf}$ as the low energy cut-off energy 
of the scaling.~\cite{Schmidt.10} Although this energy scale  --- being related to the lifetime of a slave particle --- 
has no direct physical meaning, neverheless, it  appears naturally in the 
FRG scheme (see Eq.~\eqref{eq:cut_off_function}), and is directly related to the spin relaxation rate. 
The rate $\Gamma_{\rm pf}$  can be most easily defined as the imaginary part of the 
retarded pseudofermion self-energy, which  can also be expressed in terms of the bigger and lesser 
pseudofermion self-energies as 
\be
\Gamma_{\rm {pf}} =
{i\over 2}\sum_s  \lim_{\omega\to \lambda_s}\left (\Sigma_{s}^>(\omega)- \Sigma_{s}^<(\omega)\right ).
\ee

\begin{figure}[t]
\includegraphics[width=0.8\columnwidth,clip]{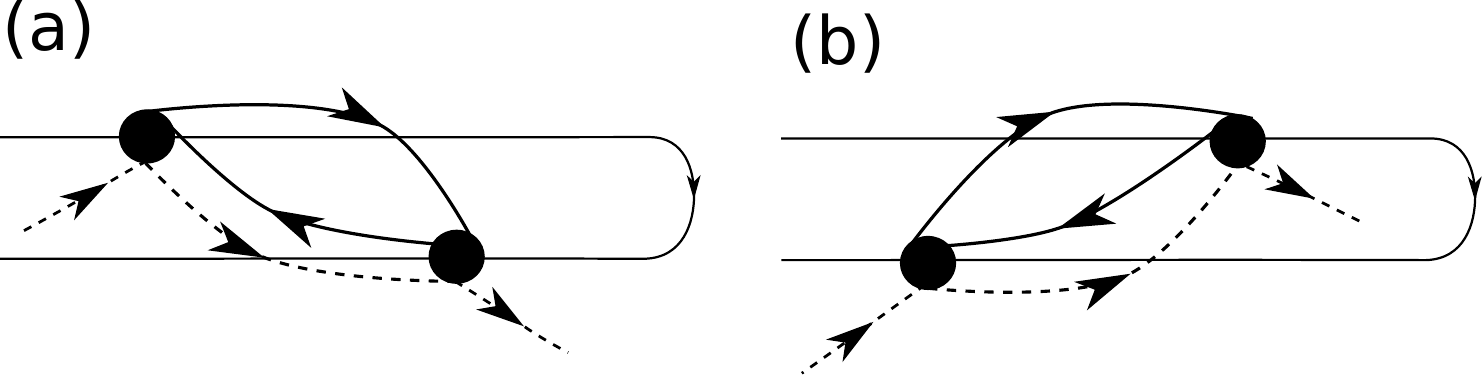}
\caption{Second order diagrams for the Keldysh components of the self-energy. In panel (a)
we represent $\Sigma^{>}(t)$, while in  panel (b) $\Sigma^{<}(t)$ is diplayed.}
\label{fig:self_energy}
\end{figure}
The  second order FRG diagrams for  $\Sigma^{\lessgtr}_{s}$ are  shown 
in Fig.~\ref{fig:self_energy}.
The leading order (perturbative) expression of $\Gamma_{\rm pf}$ can be obtained by using the bare 
exchange couplings, 
\begin{equation}
\Gamma^{(0)}_{\rm pf} = {\pi\over 8}\sum_{\al ,\beta}\sum_{i, s, s'}\, \int \rmd \omega\,
|S^i_{ss'}|^2 | j_{\alpha\beta}|^2\, \bar f_{\beta}(\omega)
 \, f_{\alpha}(\omega-\lambda_{ss'}).
\label{eq:pf_decoherence}
\end{equation}
Evaluating Eq.~\eqref{eq:pf_decoherence} in the asymptotic limit, when 
again one of the variables $T, V, B$ is much larger that the other ones, we find that  
$\Gamma^{(0)}_{\rm pf } = (3/8)\; \Gamma^{(0)}_K$, with $\Gamma^{(0)}_K$ the leading order expression of the Korringa rate, 
 Eq.~\eqref{eq:Korringa}. 

The complete  expression of the running rate, $\Gamma_{\rm pf}(a)$ is somewhat involved, 
and can be expressed as 
\begin{multline}
\Gamma_{\rm pf}(a) = \frac \pi  2 \sum_{\alpha\beta\s\s'}\sum_{s s'}
\int{\rm d}\omega\, g_{\alpha\beta}^{\s s; \s's'}(a, \omega-\frac{\lambda_{ss'}}{2})\times\\
\bar f_{\beta}(\omega)\, 
g_{\beta\alpha}^{\s' s'; \s s}(a, \omega-\frac{\lambda_{ss'}}{2})\,
f_\alpha(\omega-\lambda_{ss'})
\end{multline}
Similar to $\Gamma_K$, the renormalized rate $\Gamma_{\rm pf}$ can be obtained by solving the RG equations 
self-consistently (now using $\Gamma_{\rm pf}(a)$ as a cut-off), and then taking the $a\to 0$ limit.

It is a delicate problem by itself to decide which  of the  two rates, $\Gamma_K$ and $\Gamma_{\rm pf}$ 
should be used as an infrared  cut-off.  This  issue has been   discussed  in detail 
in Ref.~\onlinecite{Paaske.04}, by doing pertubative calculations up to $3^{rd}$ order in $j$. This analysis shows that  
in the expression for the conduction electrons T-matrix, e.g., several   logarithmic singularities 
emerge which are cut off by different rates.  
Since $\Gamma_K$ and $\Gamma_{\rm pf}$ differ only by a numerical prefactor of order 1, 
here we shall not distinguish  them and we choose  to use 
the physical spin relaxation rate $\Gamma_K$  as an infrared cut-off, similar to
Ref.~\onlinecite{Rosch.05}.

\section{Finite-frequency noise}
\label{sec:noise}

As mentioned earlier, the current operator, $\hat I_\alpha (t)$, does not commute with itself at different times, and therefore, several different current-current correlators can be defined
(see  Sec.~\ref{sec:current_noise}).
 Emission and absorption noise measurements, e.g.,  give access to the "bigger" correlation function, 
\begin{equation}
S_{\alpha\beta}^> (\omega) = \int_{-\infty}^{\infty} {\rm d} t \; e^{i \omega t} S_{\alpha\beta}^> (t)
\end{equation}
with $S_{\alpha\beta}^>(t)$ defined in Eq.~\eqref{eq:up-noise}. 
The spectrum  $S_{\alpha\beta}^>(\omega)$ is in general a complex, not symmetrical function in frequency.  
As discussed   in Ref.~\onlinecite{Gavish.00},  
$S^> (\omega)$ can be interpreted as the rate by which the system 
absorbs ($\omega >0$) or emits ($\omega<0$) photons of energy $|\hbar \omega|$. 
While usual amplifiers measure a combination of emission and absorption 
processes, using a \emph{quantum detector} gives the opportunity to measure separately the emission and absorption noise. Depending on whether photons are emitted or absorbed by the 
quantum detector,  one can thus measure the  $\omega<0$ (emission) or the
$\omega>0$ (absorption) part of $S^> (\omega)$,~\cite{Clerk.10}
\begin{equation}
S^{\rm{em/ab}}_{\alpha\beta}(\omega>0)\equiv
S^{<}_{\alpha\beta}(\pm \omega). 
\end{equation}

%
%

Time dependent current-current correlation functions may depend not only  
on the way noise is measured, but also on the precise 
spatial location (electrode) where currents are measured.  
In our case, however, current conservation [Eq.~\eqref{eq:current_operator}]
 guarantees that it is enough to focus only on one noise component, 
say $S_{LL}^{\gtrless}$,  as all the other ones are trivially related to it,
\begin{equation}
S^{\gtrless}_{RR}(\omega) =-S^{\gtrless}_{LR}(\omega)= S^{\gtrless}_{LL}(\omega)\,. 
\end{equation}
We shall therefore focus on $S_{LL}^{\gtrless}$ in what follows.

\begin{figure}[t]
\includegraphics[width=\columnwidth,clip]{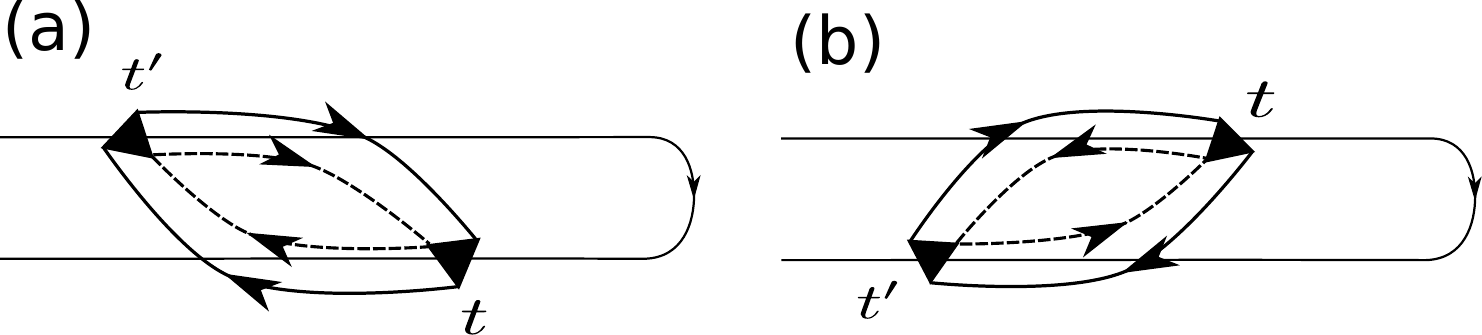}
\caption{ Diagrams for the  noise components.  $S^{>}(t, t')$ is represented in panel (a), and $S^{<}(t,t')$ in panel (b).}
\label{fig:noise}
\end{figure}

\begin{figure*}[t]
\begin{center}
\includegraphics[width=1.2\columnwidth,clip]{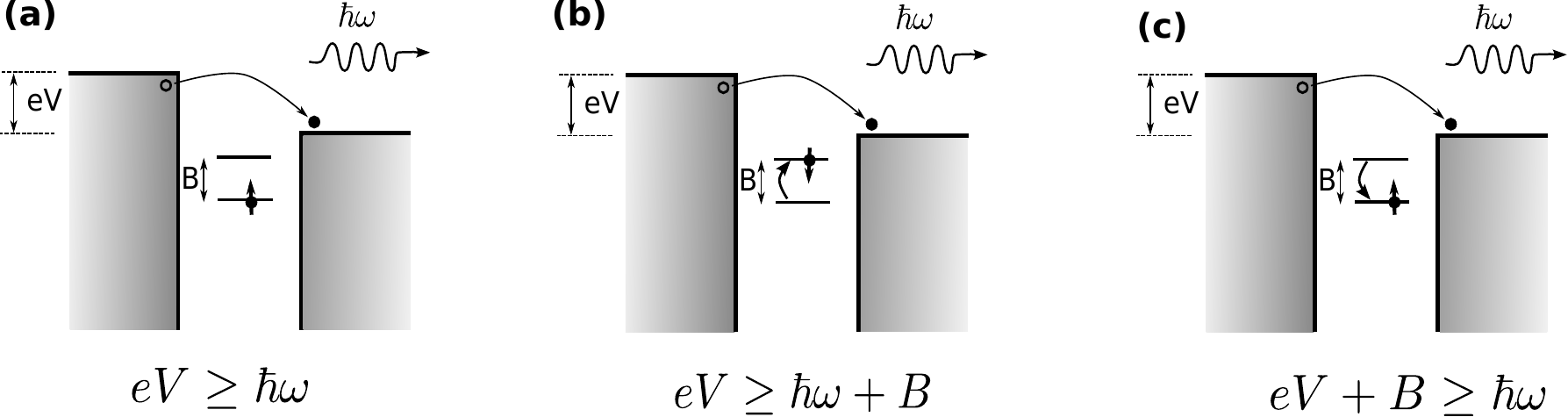}
\caption{ Basic  photon emission processes.}
\label{fig:processes}
\end{center}
\end{figure*}

A real time  diagrammatic representation of the noise correlation functions, Eqs.~\eqref{eq:up-noise}  and \eqref{eq:low-noise} is presented in Fig.~\ref{fig:noise}. Using 
Eq.~\eqref{eq:nonlocal_current}, the spectra $S_{LL}^{\gtrless}$  can be 
obtained via Fourier transformation   
\bea 
S_{LL}^{>}(\omega) &=& \frac{e^2}{16}\;
\sum_{ss'} p_s \sum_{\alpha\alpha'\sigma \sigma'}
\int \rmd\tilde \omega\;
 V_{\alpha\alpha'}^{ \sigma s; \sigma' s'}(\tilde\omega_-,\tilde\omega_+)
\nonumber
\\
&&  G_{\alpha'}^{>}(\tilde\omega_+)\,
V_{\alpha ' \alpha}^{ \sigma ' s' ; \sigma s}(\tilde\omega_+,\tilde\omega_-)\, 
G_{\alpha}^{<}(\tilde\omega_-)
\;,\\ &&\nonumber\\
S_{LL}^{<}(\omega) &=& \frac{e^2}{16}\;
\sum_{ss'}p_s \sum_{\alpha\alpha'\sigma \sigma'}
\int \rmd\tilde \omega\;
 V_{\alpha\alpha'}^{ \sigma s; \sigma' s'}(\tilde\omega_-,\tilde\omega_+)
\nonumber
\\
&&  G_{\alpha'}^{<}(\tilde\omega_+)\,
V_{\alpha ' \alpha}^{ \sigma ' s' ; \sigma s}(\tilde\omega_+,\tilde\omega_-)\,
G_{\alpha}^{>}(\tilde\omega_-)
\;,
\eea
with $\tilde\omega_\pm=\tilde \omega \pm \omega/2 \pm \lambda_{ss'}/2$,  
$G^{\gtrless}_\alpha(\omega)$ the electronic Green's functions,   and $p_s$ the probability of  
the spin (pseudofermion) being in state $s$. These latter  are computed self-consistently   
in terms of the transition rates, $\Gamma_s$, by solving the 
detailed balance equation, 
$p_\Uparrow \Gamma_{\Downarrow} =p_{\Downarrow}\Gamma_{\Uparrow} $.

\begin{figure}[b]
\includegraphics[width=0.8\columnwidth,clip]{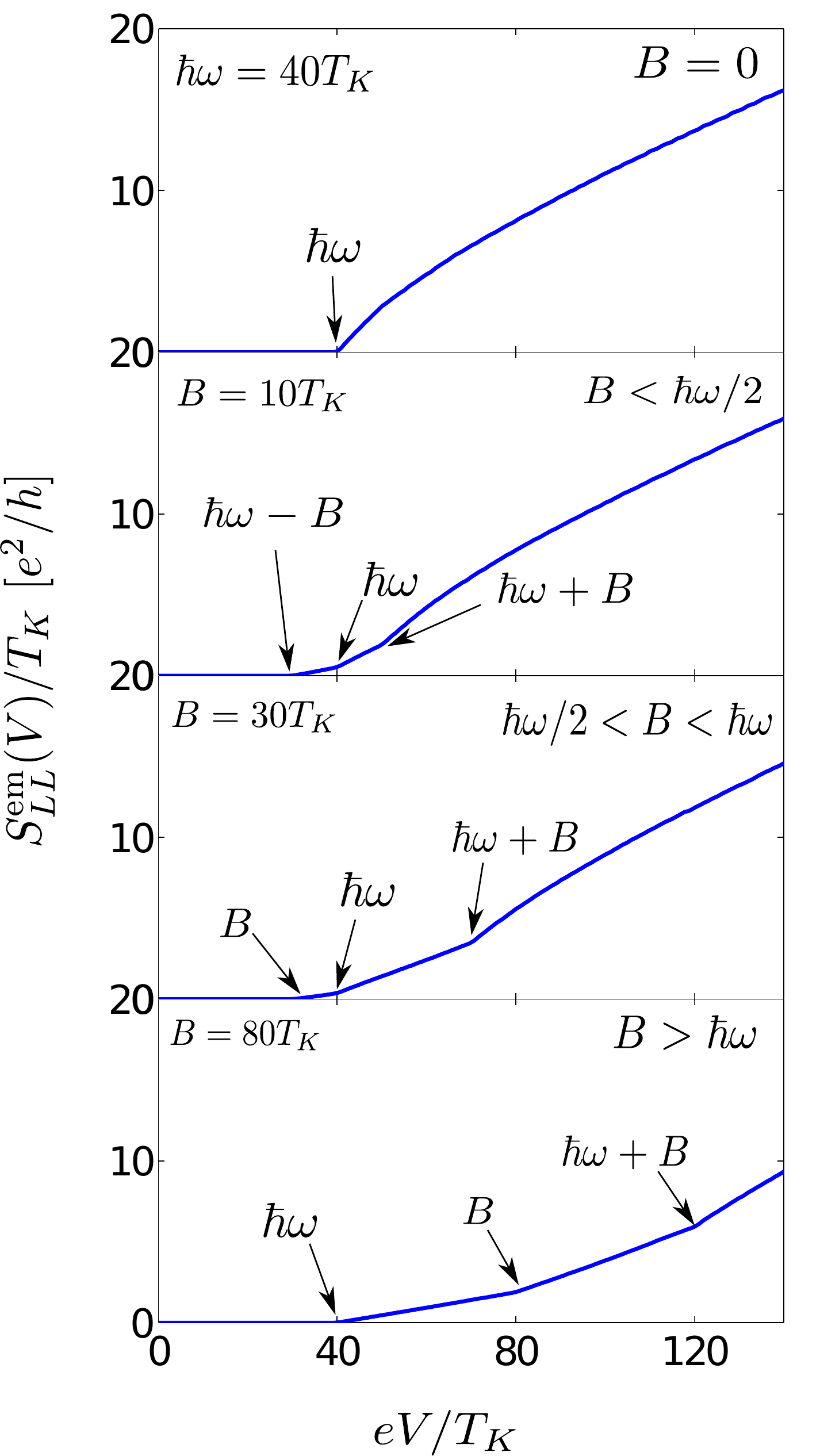}
\caption{ (color online)
Voltage dependence of the emission noise $S^{\rm em}_{LL} (\omega)$ for
a fixed frequency, $\omega = 40 \,T_K$, and for different values of the magnetic field.
The arrows indicate the positions of the kinks in the spectra.
}
 \label{fig:emission_noise}
\end{figure}

\subsection{Emission noise spectra}

In the experiments of Ref.~\onlinecite{Basset.12} one measures the emission noise of a 
nano-circuit at a fixed frequency, $\omega$, as a function of the bias voltage, 
$S^{\rm em} (\omega, V) $. 
In Fig.~\ref{fig:emission_noise} we therefore display the zero-temperature voltage dependence of the QD's emission noise  at a fixed finite frequency $\omega = 40\,T_K$  for various external magnetic fields.  The spectra develop kinks (appearing as sharp steps in ${\rm d}S^{\rm em}/{\rm d} V$, as shown in the introduction), associated with the opening of various new photon-emission channels, displayed in Fig.~\ref{fig:processes}. These kinks can be understood as follows.

In the absence of an external field, $B=0$, there is a single kink (threshold), located at $\hbar \omega/e$. Below this voltage the energy gain of an electron passing through the circuit is not enough to trigger photon emission.  Above this threshold, on the other hand, 
the emission noise exhibits  a sharp increase 
followed by a less steep, close to linear dependence at larger voltages, $V\gg T_K$. The sharp increase close to threshold is a manifestation of the non-equilibrium Kondo effect,~\cite{Basset.12} and amounts  to a a peak in the ${\rm d}S^{\rm em}/{\rm d} V$ spectrum, as also observed 
experimentally.\cite{Basset.12}

For small magnetic fields,  $B< \hbar \omega/2$ two more kinks appear at $V = \hbar \omega \pm B$ (see second panel of Fig.\ref{fig:emission_noise}). These can be understood as follows: For very small voltages the spin is polarized by the external filed ($p_\Uparrow =1$, $p_\Downarrow =0$). Once the voltage becomes larger than the splitting of the two spin states, $B$, spin flip processes can 
populate the state $|\Downarrow\rangle$, and  $p_\Downarrow$ becomes finite. 
Therefore, photon emission becomes possible through spin flip processes (shown in Fig.~\ref{fig:processes}),  once the voltage reaches the threshold  $eV = \hbar \omega- B$. Conversely, 
a new spin flip scattering channel opens at $eV = \hbar \omega + B$, where the potential energy gain of an electron passing through the QD is converted to a spin excitation and 
the energy of the emitted photon (see Fig.~\ref{fig:processes}.b). 

The situation explained in the previous paragraph changes slightly, once the magnetic field becomes somewhat larger that $\hbar \omega/2$ (see Fig.~\ref{fig:processes}.c). 
In this case $p_\Downarrow$ remains zero as 
long as $eV< B$. However, spin flip emission becomes energetically possible immediately once
the voltage reaches $B$ and thus $p_\Downarrow$ becomes non-zero, since in this case the voltage bias automatically satisfies the condition $eV> B> \hbar \omega -B$. Correspondingly, we recover three kinks at $eV = B$, $eV = \hbar \omega$, and $eV = \hbar \omega +B$, the 
latter kink corresponding the simultaneous photon emission and 
$\Uparrow \;\to\; \Downarrow$ spin flip  process. 
Finally, for $B>\hbar\omega$ the location of the kinks remains the same as for $\hbar\omega/2<B<\hbar \omega$, but in this case emission starts at the "optical gap", $\hbar \omega$.

Though the features discussed so far seem to be relatively weak, 
experimentally one has access to 
the differential noise spectrum, i.e. to the derivative of the current noise with respect to 
the voltage, ${\rm d}S^{\rm em}(V)/{\rm d} V$.~\cite{Basset.12} This quantity, 
already presented in Fig.~\ref{fig:emission_noise_derivative} 
 displays much sharper  features than $S^{\rm em}(V)$ itself at every threshold, and should allow to identify each process unambiguously.  

\subsection{Frequency dependence of  $S^>(\omega)$ and $S^s(\omega)$}
\begin{figure}[b]
\includegraphics[width=0.8\columnwidth,clip]{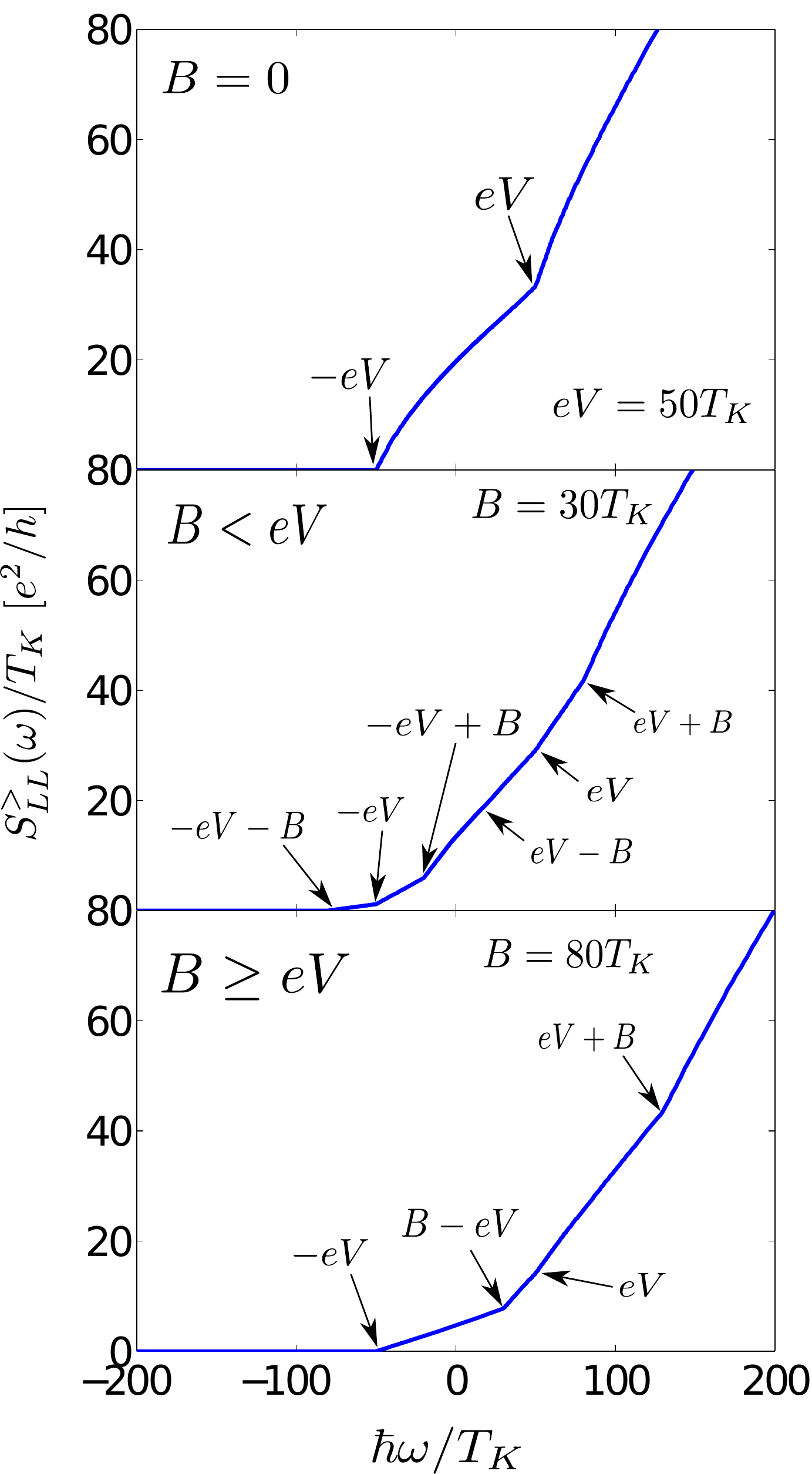}
\caption{ (color online)
Frequency dependence of the bigger noise $S^{>}_{LL} (\omega)$ at $T=0$ 
and for different values of the magnetic field. The arrows indicate the corresponding 
frequency at which weak logarithmic singularities emerge in the spectra.}
 \label{fig:bigger_noise_frequency}
\end{figure}

So far, we only discussed the behavior of $S^>(\omega)$ at a fixed negative frequency
(emission noise), as a function of external voltage. The spectrum $S^>(\omega)$ at a finite 
and fixed \emph{voltage} contains, however  more  information since it accounts   both for absorption and for emission processes. The function $S^>(\omega)$ 
is displayed in  Fig.~\ref{fig:bigger_noise_frequency} for various magnetic fields.

\begin{figure}[t]
\includegraphics[width=0.8\columnwidth,clip]{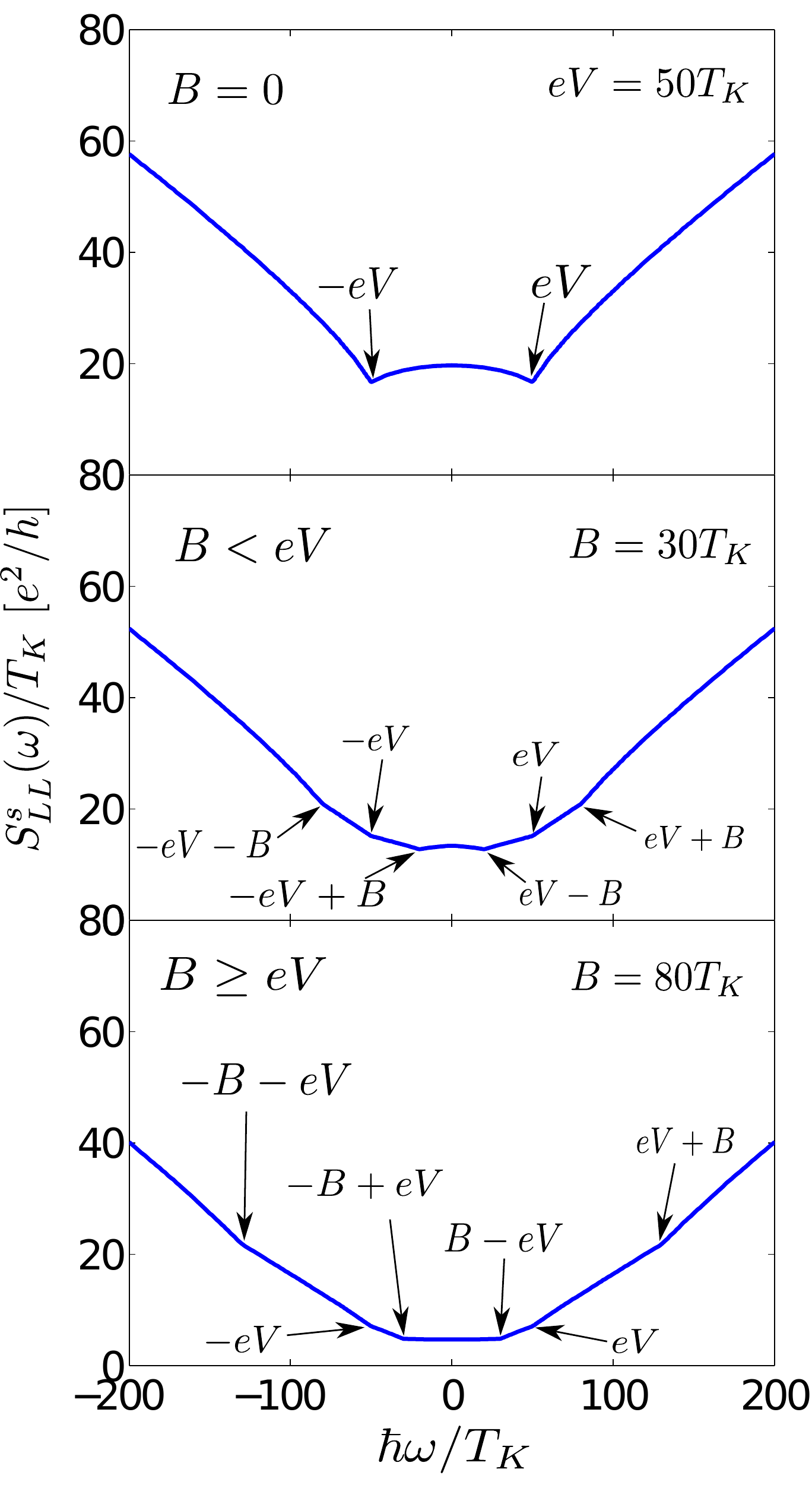}
\caption{ (color online)
Frequency dependence of the symmetric noise $S^{s}_{LL} (\omega)$ at 
$T=0$ and for different values of the external magnetic field as computed with the 
RTFRG}
 \label{fig:symmetric_noise_frequency}
\end{figure}

The  structures on the $\omega <0$ (emission) side can be understood along lines very similar to the ones presented in the previous subsection. Here, however, we need to distinguish only two regions:  For $eV< B$ the spin down state is not populated. Therefore, in this region only photons 
with energy $\hbar\omega < eV$ are emitted in a process where an electron is transferred through the QD without spin flip. Correspondingly, in this region there is only an emission threshold at $\hbar \omega = -eV$  (see third panel of Fig.~\ref{fig:bigger_noise_frequency}).
 For  $eV> B$, on the other hand, the spin 
$\Downarrow$ state of the QD gets populated, and all three emission processes of Fig.~\ref{fig:emission_noise} become active. Correspondingly a threshold is shifted to  
$\omega = -(eV+ B)$ and two more kinks appear at frequencies 
$\hbar \omega = -eV$ and $\omega = -(eV- B)$.\footnote{Throughout this discussion we assume $eV,B> 0$.} 

Understanding the $\omega >0$ (absorption)  side is much easier:  There, all three 
absorption processes are allowed (if the photon's energy is large enough), and 
correspondingly, three kinks are always recovered at frequencies 
$\hbar \omega = eV$ and $\omega = |eV\pm B|$. 

The structure of the symmetrized noise, measured by a conventional amplifier is much 
simpler: since the symmetrized noise is just a combination of the 'bigger' and 'lesser' noises
$S^s(\omega) = (S^>(\omega)  + S^<(\omega))/2 $, the kinks associated with the thresholds of the various absorption and emission  processes appear now in it symmetrically at all three frequencies, 
$\hbar \omega = \pm eV$ and $\omega = \pm (eV\pm B)$ (see Fig.~\ref{fig:symmetric_noise_frequency}).

\section{Conclusions}\label{sec:conclusions}

In this work we  have developed in detail 
 a real time functional renormalization group (FRG) formalism, originally proposed in 
Ref.~\onlinecite{Moca.11},  and 
used it to study the finite frequency noise in a quantum dot, subject to an external  
magnetic field   in the  local moment regime.
We have shown that within a systematic real time FRG formalism, similar to the interaction vertex,  the current vertex necessarily becomes non-local in time, and  a new RG equation must be 
constructed to account for the renormalization of the current vertex. The structure of our real time  RG scheme thus resolves the long-standing problem of current conservation. 
Our approach sums up all leading logarithmic contributions, and is valid at any frequency, $\omega$, voltage $eV$, or magnetic field $B$,  provided that $\max\{\omega, eV,B \}\gg T_K$. 
As demonstrated in Ref.~\cite{Basset.12}, the present theory accounts well for the features 
observed experimentally.

We have solved the FRG equations in Fourier space numerically, and computed 
the emission/absorption  and symmetrized  noise spectra through a voltage biased QD. 
   A very rich behavior is found. 
In the differential emission noise of the QD
${\rm d}S^{\rm em}(\omega,V)/{\rm d}V$ (measured at a finite frequency $\omega$), 
logarithmic singularities appear at the thresholds, $\hbar \omega =  e V$ and  
$\hbar \omega =  |e V\pm B|$, corresponding to the opening of spin-conserving and 
spin-flip  emission channels, and reflecting the presence of the non-equilibrium Kondo effect. 
The experimentally measured peaks (anomalies) in the differential emission spectra
of Ref.~\onlinecite{Basset.12},  ${\rm d}S^{\rm em}/{\rm d}V$ are thus predicted  to split up in a magnetic field into two or three singular features (steps), as  
shown in Fig.~\ref{fig:emission_noise_derivative}. 
These results  agree  in large with the ones
presented in Ref.~\onlinecite{Muller.13}, where similar quantities were investigated by 
using a  somewhat different (and more involved) technique, formulated in terms of a Liouvillian approach 
on the Keldysh contour.  
Though the results (locations and general structure of $\mathrm dS/\mathrm dV $ anomalies, etc.) of the two approaches are rather similar, there are, however, some differences, too, worth mentioning. Maybe the most important difference between the method of Ref.~\onlinecite{Muller.13}
and ours  is the way the two formalisms treat spins and spin  relaxation.  While our approach is based upon a pseudofermion formalism, the computations of Ref.~\onlinecite{Muller.13} are carried out directly in terms of the impurity spin. 
 A great advantage of the pseudofermion approach 
discussed here is that it allows a systematic and relatively easy  computation of the dynamical  vertex function, and systematically incorporates 
dynamical logarithmic corrections. Including spin relaxation, however, is not entirely straightforward within this approach.
 In a magnetic field in the $z$ direction, e.g., the spin acquires a finite expectation value, $\langle S^z(t)\rangle\ne0$. Correspondingly, the spin-spin correlation  function $\langle S^z(t)S^z(0)\rangle$  does not decay to zero, and its  Fourier transform therefore contains a delta peak at $\omega=0$. 
As argued in Ref.~\onlinecite{Glazman.05}, in equilibrium, this  amounts in the appearance of purely elastic  scattering processes in a magnetic field, originally absent   for $B=0$. At a finite bias, if indeed still present,   such  elastic left-right charge transfer processes could give rise to a sharp step in the $\mathrm dS/\mathrm dV$ curves at $\omega =eV$.
Indeed, such a sharp step was found at $\omega =eV$ within the approach of Ref.~\onlinecite{Muller.13} at $T=0$ 
temperature, while the other steps were found to be washed out due to spin relaxation.

Reproducing the previously-mentioned finite step -- if it indeed exists --  is far from trivial within the pseudofermion approach.  
In its simpler form (where certain vertex corrections are neglected), the pseudofermion method incorporates spin  relaxation only through the pseudofermion's lifetime, and, correspondingly, it predicts a broadened resonance even at 
$\omega=eV$.  In Figs.~Ê\ref{fig:emission_noise}--\ref{fig:symmetric_noise_frequency}, for simplicity, we neglected the pseudofermion's lifetime 
within the pseudofermion loop of Fig.~\ref{fig:noise} and approximated it by a non-decaying spin relaxation function. As shown in Fig.~\ref{fig:emission_noise_derivative}, incorporating the pseudofermion's 
self-consistently determined lifetime in this diagram  gives a small, but finite  width to all steps in the $\mathrm dS/\mathrm dV$ curves.
One could, of course,  replace this loop - somewhat heuristically - by a resumed pseudofermion ladder series (and thereby reproduce the non-decaying part of the  spin-spin correlation function), but we preferred to present here a self-consistent framework. 
Whether the finite jump at $\omega =eV$ -- obtained within a perturbative approach of Ref.~\onlinecite{Muller.13} -- 
indeed survives in a biased system is a rather  non-trivial, intriguing question. Observing it seems to be, unfortunately, beyond current experimental 
resolution. 

Finally, we should emphasize that the method presented here   is not only relatively easy, but also quite general. It is not just restricted to  a QD,  
but can be used for \emph{any} system with some localized degrees of freedom, coupled to conduction electrons/leads via a Kondo-like coupling, Eq.~\eqref{eq:H_int_gen}. It is thus straightforward to apply it to
 molecular singlet triplet transitions,\cite{Paaske.08}  
double quantum dots systems\cite{Makarovski.07, Lim.11, Zazunov.10}, or side-coupled molecules,\cite{Pauly.08} and 
a variety of strongly correlated nanostructures.

\section*{Acknowledgments}
We would like to thank S. Andergassen, J. Basset, H. Bouchiat, R. Deblock, M. Pletyukhov, 
and H. Sch\" oller for interesting discussions. 
This research has been supported by the French-Roumanian grant DYMESYS ( 
ANR 2011-IS04-001-01 and PN-II-ID-JRP-2011-1) 
and by the Hungarian Research Funds under grant Nos. K105149, CNK80991. 
CHC acknowledges the support from NSC grant No. 98-2918-I-009-06,
No. 98-2112-M-009-010-MY3, No. 101-2628-M-009-001-MY3
the NCTU-CTS, the MOE-ATU program,
the NCTS of Taiwan, R.O.C..


\appendix
\section{Green's functions}
\label{app:Greens_functions}
Evaluation of the path integrals gives automatically products of operators ordered along the Keldysh contour. 
Correspondingly, the average $-i \langle \psi^{(2)}_{\alpha\sigma}(t)\bar\psi^{(1)}_{\alpha'\sigma'}(t')\rangle_{\cal S}$, e.g., yields the operator 
product $-i \langle \psi_{\alpha\sigma}(t)\bar\psi^\dagger_{\alpha'\sigma'}(t')\rangle = G^>_{\alpha\sigma;\alpha'\sigma'}(t-t')$.
Using the representation $\psi_{\alpha\sigma}(t) = \int \mathrm{d}\xi c_{\alpha\sigma} (\xi)\, e^{-it (\xi + \mu_\alpha)} e^{-a |\xi|/2}$ this immediately 
yields, e.g.
\bea
&& G^{(21)}_{\alpha\sigma;\alpha'\sigma'}(t)= G^>_{\alpha\sigma;\alpha'\sigma'}(t) 
\nonumber \\
\phantom{nnn} &=&  -i  \delta_{\alpha\alpha'}
\delta_{\sigma\sigma'} e^{-i  \mu_\alpha t} 
\int \mathrm{d}\xi\, e^{-a |\xi|}\, (1-f(\xi)) \, e^{-i\xi t}\,. 
\nonumber
\eea
The itegral can be carried out at  $T=0$ temperature and yields a propagator 
$\sim -e^{-i\mu_\alpha t} /(t-ia)$. The other Keldysh propagators can be determined similarly.
They are all diagonal in the spin and lead labels, $G^{(\kappa\kappa')}_{\alpha\sigma,\alpha'\sigma'}(t)  =  \delta_{\alpha\alpha}  \delta_{\sigma\sigma'}G^{(\kappa\kappa')}_\alpha(t)  $,  and are given at $T=0$ 
temperature by
\bea
G_\alpha^{(11)} &=& G_\alpha^{t}(t)= - \frac{  e^{-i \mu_\alpha t} }  
{t- i a\; {\rm sgn}(t)}
\nonumber
\,,
\\
G_\alpha^{(22)} &=& G_\alpha^{\overline t}(t)= - \frac{e^{-i \mu_\alpha t} }  
{t +  i a\; {\rm sgn}(t)}
\nonumber
\,,
\\
G_\alpha^{(21)} &=&
G_\alpha^>(t)= - \frac{   e^{-i \mu_\alpha t}}  {t-i a} 
\nonumber \, ,
\\
G_\alpha^{(12)} &=&
G_\alpha^<(t)= - \frac{   e^{-i \mu_\alpha t}}  {t+i a} 
\nonumber \, .
\eea

The Abrikosov pseudofermion Green functions 
are diagonal, $F^{(\kappa\kappa')}_{ss'} (t) = \delta_{ss'} \, F^{(\kappa\kappa')}_{s} (t) $, and  
can be computed similar to the conduction electron propagators, by recurring to the  operator representation. They are given by the following expressions, 
\begin{eqnarray}
F^{(11)}_s(t) &=& 
F_s^t(t)=-i\, e^{-i\lambda_s t} \bigl( \Theta(t) - e^{-\beta\lambda_s } \bar \Theta(t)\bigr)  
+\dots,
\nonumber\\
F^{(22)}_s(t) &=& F_s^{\bar t}(t)= -i \,e^{-i\lambda_s t} 
 \bigl(\bar  \Theta(t) - e^{-\beta\lambda_s }  \Theta(t)\bigr)  +\dots ,
\nonumber
\\
F^{(21)}_s(t) &=&  F_s^>(t)=-i \,e^{-i\lambda_s t} + \dots , \nonumber\\
F^{(12)}_s(t) &=&  F_s^<(t)= i \, e^{-i\lambda_s t}e^{-\beta\lambda_s} + \dots ,
\nonumber
\end{eqnarray}
with $\Theta(t)$ the Heaviside function, $\bar\Theta(t)=1-\Theta(t)$, and $\beta=1/(k_BT)$.
The dots indicate subleading corrections  in $e^{-\beta\lambda_s}$, which can be dropped within the physical subspace $\sum _sf^\dagger_sf_s =1$.Ê


\bibliography{bibliography}

\end{document}